\documentclass[10pt]{llncs}

\usepackage{amsmath,amssymb,mathtools}

\usepackage[dvipsnames]{xcolor}
\usepackage[normalem]{ulem}
\usepackage{hyperref}
\hypersetup{
  colorlinks = true,
  urlcolor   = blue,
  linkcolor  = RoyalBlue,
  citecolor  = red
}

\usepackage{cite}

\usepackage{algorithm, algpseudocode}
\algnewcommand{\myAnd}{\textbf{ and }}

\newcommand{\Z}{\mathbb{Z}}
\newcommand{\F}{\mathbb{F}}
\newcommand{\Fq}{\F_q}
\newcommand{\Fm}{\F_{q^m}}

\newcommand{\Fmn}{\F_{q^m}^n}

\newcommand{\mC}{\mathcal{C}}
\newcommand{\mO}{\mathcal{O}}

\newcommand{\abf}{\mathbf{a}}
\newcommand{\bbf}{\mathbf{b}}
\newcommand{\cbf}{\mathbf{c}}
\newcommand{\ebf}{\mathbf{e}}
\newcommand{\mbf}{\mathbf{m}}
\newcommand{\nbf}{\mathbf{n}}
\newcommand{\ubf}{\mathbf{u}}
\newcommand{\tbf}{\mathbf{t}}
\newcommand{\vbf}{\mathbf{v}}
\newcommand{\wbf}{\mathbf{w}}

\DeclareMathOperator{\Gal}{Gal}
\DeclareMathOperator{\rk}{rk}
\DeclareMathOperator{\srk}{srk}
\DeclareMathOperator{\rowspan}{rowsp}
\DeclareMathOperator{\Id}{Id}
\DeclareMathOperator{\Moore}{Moore}
\DeclareMathOperator{\GL}{GL}
\DeclareMathOperator{\Tr}{Tr}
\DeclareMathOperator{\id}{id}
\DeclareMathOperator{\ev}{ev}
\DeclareMathOperator{\Gab}{Gab}
\DeclareMathOperator{\LRS}{LRS}
\DeclareMathOperator{\GLRS}{GLRS}
\DeclareMathOperator{\Short}{Short}

\newcommand{\ore}{\Fm[x;\sigma]}
\newcommand{\st}{\,:\,}

\renewcommand{\leq}{\leqslant}
\renewcommand{\geq}{\geqslant}

\begin{document}

\title{Full Structure Recovery of Linearized Reed–Solomon Codes}
\author{Valentina Astore\inst{1,2,3}\thanks{V.~A. is funded by AID (French \emph{Agence de l'innovation de défense}).} \and Martino Borello\inst{3} \and Alain Couvreur\inst{1,2}}
\institute{Inria, France. \and LIX, École polytechnique, Institut Polytechnique de Paris, France. \and Universit\'e Paris 8, Laboratoire de G\'eom\'etrie, Analyse et Applications, LAGA, Universit\'e Sorbonne Paris Nord, CNRS, UMR 7539, France. \email{valentina.astore@inria.fr } \email{martino.borello@univ-paris8.fr } \email{alain.couvreur@inria.fr}}

\maketitle

\begin{abstract}
    Linearized Reed--Solomon codes (LRS) are optimal sum-rank-metric codes that generalize both Reed--Solomon and Gabidulin codes, and benefit from excellent decoding capabilities, making them potentially interesting candidates for post-quantum code-based cryptography. Previous works have investigated the underlying geometric structure of LRS codes, identifying geometric invariants that distinguish them from random codes. Other works have proposed approaches to recovering the LRS structure from an arbitrary generator matrix, together with additional data. However, both the distinguishers and the recovery techniques proposed so far are either not fully general, or have a prohibitive computational complexity. In this paper, we propose a polynomial-time procedure for the full structure recovery of LRS codes from the sole knowledge of an arbitrary generator matrix. Our results show that, like their Hamming and rank metric counterparts, LRS codes cannot be used as they are for McEliece-like encryption. Our approach combines Overbeck-type attacks with suitable matrix manipulations and shortening techniques to recover the evaluation parameters across different ranges of code dimensions.
\end{abstract}


\section*{Introduction}
Code-based cryptography emerged at the end of the last century as a promising approach to building cryptographic schemes capable of resisting both classical and quantum attacks. In response to the rapid recent advances in quantum computing, NIST has standardized the first post-quantum schemes and has selected the code-based KEM HQC for standardization~\cite{nist2025hqc}, making code-based cryptography a central area of research in the field.

The first code-based cryptographic scheme dates back to 1978. It was introduced by McEliece and based on Goppa codes~\cite{mceliece1978public}.
A longstanding issue was that McEliece's original proposal was affected by the drawback of requiring large key sizes.
Later, Niederreiter introduced a dual formulation of the scheme and proposed instantiating it with generalized Reed--Solomon (GRS) codes~\cite{niederreiter1986knapsack}. The higher decoding capacity of GRS codes compared to Goppa ones led to a significant reduction of the key size for a similar claimed security level. However, the strong algebraic structure of GRS codes turns out to be difficult to mask and, in 1992, Sidelnikov and Shestakov proposed a devastating attack on McEliece instantiated with GRS codes~\cite{sidelnikov1992insecurity}. Subsequent proposals attempted to retain the advantages of structured codes while preventing the recovery of their underlying algebraic structure. In particular, Berger and Loidreau proposed a variant based on subcodes of GRS codes~\cite{berger2005mask}, which was subsequently broken by Wieschebrink~\cite{wieschebrink2010cryptanalysis}. Other variants of McEliece attempting to use generalized Reed--Solomon codes with an adequate masking such as \cite{W06,BBCRS14,W16,W17,KRW21,BBPR18} were all subject to efficient attacks \cite{couvreur2014distinguisher,CLT19,CL22,CPTZ25}. Similarly, McEliece variants based on algebraic geometry codes and their subcodes proposed in \cite{JM96} were shown to be vulnerable to polynomial-time structural attacks~\cite{couvreur2017cryptanalysis}. Other proposals relied on structured alternant codes. However, several of these variants were also shown to be vulnerable to structural attacks~\cite{faugere2010algebraic,otmani2010cryptanalysis,BC18}.

Another approach to obtain more compact keys has been to consider rank-metric codes instead of Hamming-metric ones. In 1991 the first rank-metric McEliece-like scheme (GPT) was proposed. It builds on Gabidulin codes~\cite{gabidulin1985theory}, the rank-metric analogues of Reed--Solomon codes. Because of the strong algebraic structure of these codes, the original GPT scheme and most of its variants \cite{FL05,G08,GRH09,RGH10,loidreau2017new,PRW18,LLP20,RPW21} were (at least partially) broken, mainly by the attacks of Gibson~\cite{gibson1995severely,gibson1996security}, Overbeck~\cite{overbeck2008distinguisher} and their extensions \cite{GOT18,OTN18,coggia2020security,BC21,BC22,CouvreurZappatore}.

Linearized Reed--Solomon codes are optimal sum-rank-metric codes that generalize both Reed--Solomon and Gabidulin codes, making them natural candidates for building post-quantum code-based primitives~\cite{dissertation,nouetowa2026post}. They were introduced in~\cite{martinezpenas2018skew} as a special class of evaluation codes, whose definition can therefore be reduced to the choice of suitable evaluation parameters. Since the security of code-based schemes typically relies on the difficulty of identifying the algebraic structure of the underlying code, the problem of distinguishing linearized Reed--Solomon codes from random ones has recently attracted considerable attention. An Overbeck-type distinguisher for (generalized) linearized Reed--Solomon codes was first proposed in~\cite{hormann2026distinguishers}. More recently, \cite{astore2026qanalogue} addressed the distinguishability problem for these codes in a broader setting. However, although these methods can identify a target code as a linearized Reed--Solomon one, none of them provides a recovery procedure for its evaluation parameters from the sole knowledge of an arbitrary generator matrix. Recovering such parameters is of particular interest since, when they are known, existing decoding algorithms allow one to efficiently decode linearized Reed--Solomon codes up to half their minimum distance. Thus, structure recovery could potentially turn a structural distinguisher into an effective decoding attack. A first partial attempt in this direction was proposed in~\cite{hormann2022distinguishing}. However, this strategy is not fully general, as it requires partial knowledge of the code's evaluation parameters. In particular, an attacker with access only to a masked generator matrix must first guess some of the code's parameters to carry out the attack, introducing an additional exponential computational cost.

\subsection*{Our contribution}
In this paper, we fill in this gap by proposing a polynomial-time procedure for a full structure recovery of the evaluation parameters of linearized Reed--Solomon codes, starting from a public masked generator matrix. First, we show that, for codes over the finite field $\mathbb{F}_{q^m}$ with dimension $k<m$, an Overbeck-type attack, combined with a suitable manipulation of the public matrix, allows one to recover the evaluation parameters of these codes. For larger dimensions, we provide a full structure recovery as long as each block of the sum-rank code has length $m$. Denoting by $N$ the length of the code and by $\omega$ the complexity exponent of linear algebra, our algorithm succeeds in recovering the evaluation parameters in  $\mO(N^{\omega+1})$ operations in $\Fm$ and $\mO(N^3)$ applications of an automorphism of $\Fm$. More fine-grained complexity formulas are given in the core of the text.

Note that our procedures address the structure recovery of linearized Reed--Solomon codes, while an apparently broader notion called \emph{generalized linearized Reed--Solomon codes} was introduced in \cite{hormann2022distinguishing}. As a byproduct, we show in Theorem~\ref{thm:LRS=GLRS} that both notions are strictly equivalent, hence focusing on linearized Reed--Solomon codes is not restrictive at all.

\section{Preliminaries}
\subsection{Context and notation}
Throughout this paper, $q$ denotes a prime power, $\Fq$ denotes a finite field with $q$ elements, and $\Fm$ denotes an extension of degree $m$. Let $\sigma$ be a generator of the Galois group $\Gal(\Fm/\Fq)$. Recall that $\sigma^m=\id_{\Fm}$ and that there exists an integer $s$ with $\gcd(s,m)=1$ such that $\sigma(a)=a^{q^s}$ for every $a\in\Fm$. Throughout, we regard $\sigma$ as fixed, although it is not assumed to be known \textit{a priori}, and omit it from the notation whenever this does not cause ambiguity. As we shall see at the end of Section~\ref{subsec:gabidulin}, $\sigma$ is the first parameter to be determined.

Vectors will be denoted with lower case bold font letters such as $\abf, \bbf$ while matrices are denoted with upper case letters $A, B$. Following the usual practice in coding theory, vectors are rows. Given a matrix $A \in \F^{u\times v}$, the row space of $A$ is denoted $\rowspan (A) \subset \F^v$.

Throughout this paper, we only consider $\Fm$-linear codes, \textit{i.e.}, a \emph{code} $\mC$ of dimension $k$ over $\Fm^n$ is a $k$-dimensional $\Fm$-linear subspace of $\Fmn$ endowed with a metric. In the following, we will in particular denote a $k$-dimensional rank-metric code over $\Fm$ as an \emph{$[n,k]_{q^m/q}$ code}. We further recall that the \emph{dual code} of $\mC$ is the code $\mC^\perp\coloneq\{\vbf \in \Fmn \st \vbf \cdot \cbf =0, \text{ for all } \cbf \in \mC\}$, where  $\cdot$ denotes the standard inner product on $\Fm^n$.

\subsection{Gabidulin codes and Overbeck's distinguisher}\label{subsec:gabidulin}
Gabidulin codes were introduced independently by Delsarte in~\cite{delsarte1978bilinear} and by Ga\-bi\-du\-lin in~\cite{gabidulin1985theory}, and later generalized in~\cite{kshevetskiy2005new}. They form a special class of rank-metric codes that can be viewed as the $q$-analogue of Reed--Solomon codes.

\begin{definition}[Evaluation vector]\label{def:evaluation_vector}
 A vector $\bbf=(b_1,\dots,b_n)\in\Fmn$ is an \emph{evaluation vector} if its components are $\Fq$-linearly independent. In particular, this forces $n\leq m$.
\end{definition}

\begin{definition}[Generalized Gabidulin code]
    Let $\bbf=(b_1,\dots,b_n)\in\Fmn$ be an evaluation vector and let $1\leq k\leq n$. The {\em generalized Gabidulin code, $\Gab_{k,\sigma}(\bbf)$, with evaluation vector $\bbf=(b_1,\dots,b_n)$ and of dimension $k$} over $\Fmn$ is the code generated by the \emph{Moore matrix}
    \[\Moore_k(\bbf)\coloneq\left(\begin{array}{ccc}
    b_{1} & \cdots & b_{n}\\
    \sigma(b_{1}) & \cdots & \sigma(b_{n}) \\
    \vdots & & \vdots \\
    \sigma^{k-1}(b_{1}) & \cdots & \sigma^{k-1}(b_{n}) \\
    \end{array}\right).\]
\end{definition}

The first distinguisher for Gabidulin codes is due to Overbeck, who used it to attack the Gabidulin-based version of the McEliece cryptosystem (GPT)~\cite{overbeck2008distinguisher}. This distinguisher exploits the dimension-growth behavior of successive \emph{$\sigma$-sums}, or \emph{$q^s$-sums}, to distinguish Gabidulin codes from random linear codes and was subsequently extended to other code-based cryptosystems and code families, see \textit{e.g.,} \cite{horlemann2018extension,GOT18,OTN18,CouvreurZappatore,coggia2020security,BC21}. We are particularly interested in its application to generalized Gabidulin codes.

Let $s$ be as above, so that $\sigma(a)=a^{q^s}$ for every $a\in\Fm$. For a rank-metric code $\mC\subseteq\Fmn$ and a positive integer $i$, we define its \emph{$i$-th $q^s$-sum} by
\[\Lambda_i^{(s)}(\mC)\coloneq\mC+\sigma(\mC)+\dots+\sigma^i(\mC).\]
The following result can be obtained by adapting~\cite[Lemma~5.1]{overbeck2008distinguisher} to generalized Gabidulin codes, see \cite[Theorem 4.8]{horlemann2017new}.
\begin{proposition} \label{prop:q_sums}
    Let $\mC=\Gab_{k,\sigma}(\bbf)\subseteq\Fmn$ be a generalized Gabidulin code of dimension $k<n$. For every positive integer $i\leq n-k$, we have
    \[\dim\big(\Lambda_i^{(s)}(\mC)\big)=k+i.\]
\end{proposition}
\noindent In contrast, the $q^s$-sums of random codes typically exhibit faster dimension growth. This behavior is known for standard Gabidulin codes~\cite[Proposition~1]{coggia2020security} and naturally extends to generalized Gabidulin codes. Thus, the dimension growth of the $q^s$-sums provides a natural analogue of the classical Overbeck distinguisher for generalized Gabidulin codes. The precise probability estimates depend on the code parameters and must be established for the chosen setting.

\medskip

Let $k<n$ and let $G$ be a generator matrix of the generalized Gabidulin code $\mC\subseteq\Fmn$. Then, an Overbeck-type attack allows one to recover an evaluation vector $\bbf\in\Fmn$ such that $\mC=\rowspan(\Moore_k(\bbf))$. The procedure is analogous to that adopted in~\cite{overbeck2008distinguisher} for standard Gabidulin codes. Note that first one has to determine the correct homomorphism $\sigma$ used to define $\mC$. This can be achieved by applying Proposition~\ref{prop:q_sums} to $\mC$ for $i=1$, \textit{i.e.}, by computing $\dim\big(\Lambda_1^{(s)}(\mC)\big)$ for each $1\leq s < m$ with $\gcd(s,m)=1$, until one finds an $s$ for which $\dim\big(\Lambda_1^{(s)}(\mC)\big)=k+1$. Note that both $s$ and $m-s$ pass this test, because $\Gab_{k,\sigma}(\bbf)=\Gab_{k,\sigma^{-1}}(\sigma^{k-1}(\bbf))$. Either choice works, so one can simply select one of the two values obtained. With the selected value of $s$, Proposition~\ref{prop:q_sums} implies that $\Lambda_{n-k-1}^{(s)}(\mC)=\Gab_{n-1,\sigma}(\bbf')$ for some $\bbf'\in\Fm^n$. Since $\Gab_{n-1,\sigma}(\bbf')$ has codimension one, an evaluation vector of $\mC$ can then easily be computed as a generator of the dual of $\Gab_{n-1,\sigma}(\bbf')$.

\subsection{Skew polynomials}
Skew polynomials were originally introduced by Ore in the more general algebraic setting of~\cite{ore1933theory}. In what follows, we recall their definition and main properties in the particular setting of finite fields. The {\em skew polynomial ring} $\ore$ is the set of polynomials over $\Fm$ equipped with standard polynomial addition and multiplication determined by the rule \[xa=\sigma(a)x, \quad a\in\Fm,\]
and extended by associativity and distributivity. The general definition of skew polynomials may also involve
a $\sigma$-derivation (see for instance \cite{martinezpenas2018skew}). In the present paper we always assume it to be zero. In particular, if $\sigma=\id_{\Fm}$, then $\ore$ reduces to the standard polynomial ring $\Fm[x]$. The {\em degree} of the non-zero skew polynomial $f=\sum_i f_ix^i$ is $\deg(f)\coloneq\max\{i\st f_i\neq0\}$, and we adopt the convention $\deg(0)\coloneq-\infty$ for the degree of the zero polynomial. We denote by ${\ore}_{<k}$ the set of skew polynomials of degree less than $k$.

\begin{remark} \label{rem:central_xm}
    Note that $\ore$ is a non-commutative polynomial ring. However, its center is $\Fq[x^m]$ (see \cite[Theorem~3]{lavrauw2012semifields}), which is isomorphic to the standard polynomial ring $\Fq[y]$. In particular, $x^m-c$ is central for all $c\in\Fq\subseteq\Fm$.
\end{remark}

Evaluating skew polynomials requires a notion of evaluation that takes their non-commutative structure into account. We introduce such a notion below.

For $a\in\Fm$ and a positive integer $i$, we say that
\[N_i(a)\coloneq\prod_{j=0}^{i-1}\sigma^j(a)\] is the {\em $i$-th partial norm} of $a$, and we adopt the convention $N_0(a)\coloneq1$. Consider the $\Fq$-linear endomorphism
\[a\sigma:\Fm\longrightarrow\Fm,\qquad b\longmapsto a\sigma(b).\]
The assignment $x\mapsto a\sigma$, together with the natural action of $\Fm$ by scalar multiplication, respects the multiplication rule defined above. Therefore, it induces a ring homomorphism from $\ore$ to $\operatorname{End}_{\Fq}(\Fm)$, where multiplication in the latter is given by composition.

In particular, for a skew polynomial $f=\sum_i f_ix^i\in\ore$, we define its evaluation at the endomorphism $a\sigma$ as
\[f(a\sigma)\coloneq \sum_i f_i(a\sigma)^{\circ i},\]
so that, for $b\in\Fm$, we obtain
\[f(a\sigma)(b)= \sum_i f_i(a\sigma)^{\circ i}(b) = \sum_i f_iN_i(a)\sigma^i(b).\]
Then, for $a,b\in\Fm$, we define the \emph{evaluation map for skew polynomials} as
\[\ev_{a,b} \st \ore \longrightarrow \Fm, \quad f \longmapsto f(a\sigma)(b).\]
Note, in particular, that $\ev_{a,b}$ is $\Fm$-linear. Moreover, for fixed $f\in\ore$ and $a\in\Fm$, the map $b\mapsto\ev_{a,b}(f)$ is $\Fq$-linear.

If $\bbf = (b_1, \dots, b_n)\in\Fm^n$, we define $\ev_{a, \bbf}$ so that 
\[\ev_{a, \bbf}(f) \coloneq (\ev_{a,b_1}(f), \dots, \ev_{a, b_n}(f)), \quad \text{for all } f\in\ore.\]
More generally, given $\abf=(a_1,\dots,a_t)\in\Fm^t$ and $\bbf=(\bbf_1,\dots,\bbf_t)\in\Fm^{\mathbf{n}}$, with $\bbf_i=(b_{i,1},\dots,$ $ b_{i,n_i})\in\Fm^{n_i}$ for $1\leq i \leq t$, we extend the evaluation map above to the multi-point evaluation map
\[\ev_{\abf,\bbf}(f) \coloneq \left(\ev_{a_i,b_{i,1}}(f),\dots,\ev_{a_i,b_{i,n_i}}(f) \right)_{1\leq i \leq t}, \quad \text{for all } f\in\ore.\]

We conclude this section with a classical statement on evaluations that can be seen as a skew polynomials analogue of Lagrange's interpolation.
\begin{lemma}\label{lem:Lagrange}
    Let $n \leq m$ and $v_1, \dots, v_n$
    be $\Fq$-linearly independent elements of $\Fm$.
    Then, for any $(w_1, \dots, w_n) \in \Fmn$, there is a unique $f\in \Fm [x; \sigma]_{<n}$ such that $(w_1, \dots, w_n) = \ev_{1, \vbf}(f) = (f(\sigma)(v_1), \dots, f(\sigma)(v_n))$.
\end{lemma}
\begin{proof} See for instance \cite[Lemma~2]{martinezpenas2017rank}
or \cite[Lemma~6]{kuijper2014list}. \qed
\end{proof}

\section{Linearized Reed--Solomon codes and problem statement} \label{sec:problem_statement}

\subsection{Linearized Reed--Solomon codes}
In this section, we recall the definition of linearized Reed--Solomon codes and give a precise formulation of the problem addressed in this paper.

Let $\nbf=(n_1,\dots,n_t)\in\Z_{>0}^t$. We set $\Fm^\nbf\coloneq\Fm^{n_1}\times\cdots\times\Fm^{n_t}$. The {\em sum-rank weight} of $\vbf=(\vbf_1,\cdots,\vbf_t)\in\Fm^\nbf$ is
\[w_{\srk}(\vbf)\coloneq\sum_{i=1}^t\rk_{\Fq}(\vbf_i),\]
where $\rk_{\Fq}(\vbf_i)$ denotes the dimension of the $\Fq$-span of the entries of $\vbf_i$. For $k\leq n_1+\dots+n_t$, an \emph{$[\nbf,k]_{q^m/q}$ sum-rank-metric code} $\mC$ is a $k$-dimensional $\Fm$-subspace of $\Fm^\nbf$, endowed with the sum-rank metric. In particular, we say that $\nbf$ is the \emph{profile} of $\mC$.

Linearized Reed--Solomon (LRS) codes are a special class of sum-rank-metric codes, which generalize both Reed--Solomon and Gabidulin codes. They were originally introduced in~\cite{martinezpenas2018skew} as evaluation codes of skew polynomials. To present them, we first recall the notion of \emph{evaluation pair}, which will provide the foundation for the definition of LRS codes.

\begin{definition}[Evaluation pair]\label{def:evaluation_pair} We say that $(\abf,\bbf)\in\Fm^t\times\Fm^\nbf$ is an {\em evaluation pair} if the norms over $\Fq$ of the components of $\abf$ are non-zero and pairwise distinct, and each component of $\bbf$ is an evaluation vector (see Definition~\ref{def:evaluation_vector}). 
\end{definition}
In particular, the conditions on evaluation pairs impose the constraints $t\leq q-1$ and $n_i\leq m$ for all $1\leq i \leq t$.

For $a\in\Fm$ and a positive integer $k$, we further define the diagonal matrix
\[D_k(a)\coloneq\left(\begin{array}{cccc}
    1 & & & \\
    & N_1(a) & & \\
    &  & \ddots & \\
    & & & N_{k-1}(a) \end{array}\right)\in\Fm^{k\times k}.\]
\begin{remark}\label{rem:diagonal_a}
    Let $a,a'\in\Fm$. Since $\sigma^m=\id_{\Fm}$, for any $i>0$, we have that
    \[\sigma^{i}(a)=\sigma^{i\!\!\mod{m}}(a) \quad \text{and} \quad N_i(a)=N_{i\!\!\mod{m}}(a)\big(N_m(a)\big)^{\lfloor\frac{i}{m}\rfloor}.\]
    Moreover, by the multiplicativity of partial norms, $D_k(a)D_k(a') = D_k(aa')$. Finally, $D_k(a)\in\GL_k(\Fm)$ if and only if $a\neq0$ and $k\geq 2$, in which case $D_k(a)^{-1}=D_k(a^{-1})$.
\end{remark}

\begin{definition}[Linearized Reed--Solomon (LRS) code]\label{def:LRS}
    Let $(\abf,\bbf)\in\Fm^t\times\Fm^\nbf$ be an evaluation pair and let $k\leq n_1+\dots+n_t$ be a positive integer. The {\em $[\nbf,k]_{q^m/q}$ linearized Reed--Solomon code with evaluation pair $(\abf,\bbf)$} (see Definition~\ref{def:evaluation_pair}) is defined as \[\LRS_k(\abf,\bbf)\coloneqq\big\{\ev_{\abf,\bbf}(f)\st f\in\ore_{<k}\big\}.\]
\end{definition}
\noindent A \emph{standard generator matrix} for $\LRS_k(\abf,\bbf)$ is the block matrix \[G_k(\abf,\bbf)\coloneq(G_1\mid\cdots\mid G_t)\]
with
\begin{align*}
    G_i &\coloneq \underbrace{\left(\begin{array}{cccc}
    1 & & & \\
    & N_1(a_i) & & \\
    &  & \ddots & \\
    & & & N_{k-1}(a_i) \end{array}\right)}_{= D_k(a_i)} \ \underbrace{\left(\begin{array}{ccc}
    b_{i,1} & \cdots & b_{i,n_i}\\
    \sigma(b_{i,1}) & \cdots & \sigma(b_{i,n_i}) \\
    \vdots & & \vdots \\
    \sigma^{k-1}(b_{i,1}) & \cdots & \sigma^{k-1}(b_{i,n_i}) \\
    \end{array}\right)}_{= \Moore_k(\bbf_i)},
\end{align*} for $1\leq i \leq t$. Note that $D_k(a_i)\in\GL_k(\Fm)$, hence $G_i$ is a generator matrix for the generalized Gabidulin code $\Gab_{k,\sigma}(\bbf_i)$ over $\Fm^{n_i}$.

We recall that, in the sum-rank-metric setting, the \emph{dual code} of an $[\nbf,k]_{q^m/q}$ code $\mC$ is an $[\nbf,(n_1+\dots+n_t)-k]_{q^m/q}$ code, defined as
\[\mC^\perp:=\{\vbf \in \Fm^{\nbf} \st \vbf\cdot \cbf=0, \text{ for all } \cbf\in \mC\}.\]
Regarding LRS codes we have the following statement.
\begin{theorem}[\!\! {\cite[Theorem~4]{martinezpenas2019reliable}}]\label{thm:LRS_duality}
With the notation of Definition~\ref{def:LRS}, there exists an evaluation pair $(\abf^\perp, \bbf^\perp)$ such that
\[{\LRS_k^{\sigma}(\abf, \bbf)}^{\perp} = \LRS_{n_1+\dots + n_t-k}^{\sigma^{-1}}(\abf^\perp, \bbf^\perp).\]
\end{theorem}
In addition, an explicit description of the evaluation pair $(\abf^\perp, \bbf^{\perp})$ is given in 
\cite[Section 2.5]{hormann2022distinguishing}.

\subsection{Generalized LRS codes and LRS codes coincide}
In~\cite[Definition~6]{hormann2022distinguishing}, the authors proposed the following generalized notion for LRS codes. Let $(\abf,\bbf)\in\Fm^t\times\Fm^\nbf$ be an evaluation pair, $\vbf=(v_1,\dots,v_t)\in(\Fm^*)^t$ and $k\leq n_1+\dots+n_t$ be a positive integer. The {\em $[\nbf,k]_{q^m/q}$ generalized linearized Reed--Solomon (GLRS) code with evaluation parameters $(\abf,\bbf)$ and $\vbf$} is defined as \[\GLRS_k(\abf,\bbf,\vbf)\coloneqq\Big\{\big( v_i \ev_{a_i,\bbf_i}(f)\big)_{1\leq i\leq t} \st f\in\ore_{<k}\Big\}.\]
One could have considered this family of codes which looks more general than the
LRS one. It turns out that
it is not.

\begin{theorem}\label{thm:LRS=GLRS}
    Let $(\abf, \bbf) \in \Fm^t \times \Fm^{\nbf}$ be an evaluation pair
    and $\vbf \in {(\Fm^{*})}^t$. Then, for any $k \leq n_1+ \cdots + n_t$,
    \[\GLRS_k(\abf, \bbf, \vbf) = \LRS_k(\abf', \bbf'),\]
    where 
    \[\abf' = (a_1 v_1 \sigma(v_1)^{-1}, \dots, a_t v_t \sigma(v_t)^{-1}) \quad \text{and} \quad \bbf' = (v_1 \bbf_1, \dots, v_t \bbf_t).\]
\end{theorem}

\begin{proof}
For any $a,b,v\in\Fm^*$, let $a'\coloneq av\sigma(v)^{-1}$ and $b'\coloneq vb$. Then, by the multiplicativity of $\sigma$ and that of the partial norms, for all $f=\sum_j f_j x^j\in\ore$,
\begin{align*}
    \ev_{a',b'}(f) &= \sum_j f_j N_j\big( a v \sigma(v)^{-1} \big) \, \sigma^j(v b) \\
    &= \sum_j f_j N_j(a) \prod_{h=0}^{j-1} \sigma^h(v) \prod_{h=0}^{j-1} \sigma^h\big(\sigma(v)^{-1}\big) \, \sigma^j(v) \sigma^j(b) \\
    &= \sum_j f_j N_j(a) \, v \prod_{h=1}^{j-1} \sigma^h(v) \prod_{h=1}^{j} (\sigma^h(v))^{-1} \, \sigma^j(v) \sigma^j(b) \\
    &= \sum_j f_j N_j(a) \, v \sigma^j(b) = v \ev_{a,b}(f).
\end{align*}
Hence, setting $\abf'\coloneq\big(a_iv_i\sigma(v_i)^{-1}\big)_{1\leq i \leq t}$ and $\bbf'\coloneq(v_i\bbf_i)_{1\leq i\leq t}$, we get
\[\LRS_k(\abf',\bbf') = \GLRS_k(\abf,\bbf,\vbf). \tag*{\qed}\]
\end{proof}

\subsection{Statement of the problem}
We now give a precise formulation of the problem addressed in this paper.
\begin{problem} \label{prob:problem_statement}
Let $2\leq t < q$, $\nbf\coloneq(n_1,\dots,n_t)\in\Z_{>0}^t$ and $2 \leq k \leq n_1+\cdots +n_t$. Given an arbitrary full-rank generator matrix $G\in \Fm^{k\times (n_1+\ldots +n_t)}$ recover, if it exists, an evaluation pair $(\abf,\bbf)$ such that 
\[\rowspan(G)=\LRS_k(\abf,\bbf).\]
\end{problem}

Note that the input matrix to Problem~\ref{prob:problem_statement} can be expressed as $G=S G_k(\tilde\abf,\tilde\bbf)$, where $(\tilde\abf,\tilde\bbf)$ is an unknown evaluation pair and $S\in\GL_k(\Fm)$ is an unknown change-of-basis matrix.
\begin{remark}
    In the usual \emph{masking} procedure underlying McEliece-type cryptosystems, a standard generator matrix is also multiplied on the right by a block monomial matrix $B\in\GL_{n_1+\dots+n_t}(\Fm)$, which is a sum-rank-metric isometry. In particular, $B$ may both scramble the columns within each block and permute the blocks themselves. However, up to a relabeling of the blocks, its effect can be absorbed into the unknown evaluation pair $(\tilde\abf, \tilde\bbf)$. Therefore we may work with the simpler description given above.
\end{remark}

Before addressing Problem~\ref{prob:problem_statement}, we establish a preliminary result that will be crucial to the development of our recovery procedure.

For $\mu\in\Fm$ and a positive integer $k$, we define the diagonal matrix
\[\Delta_k(\mu)\coloneq\left(\begin{array}{cccc}
    \mu & & & \\
    & \sigma(\mu) & & \\
    &  & \ddots & \\
    & & & \sigma^{k-1}(\mu) \end{array}\right)\in\Fm^{k\times k}.\]
\begin{remark}\label{rem:diagonal_mu}
    Clearly, $\Delta_k(\mu)\in\GL_k(\Fm)$ if and only if $\mu\neq0$, and $\Delta_k(\mu)\Delta_k(\mu')=\Delta_k(\mu\mu')$ for all $\mu,\mu'\in\Fm$. Moreover, since $\sigma$ is multiplicative and fixes $\Fq$, for every $\bbf\in\Fmn$, $\mu\in\Fm$ we have $\Moore_k(\mu\bbf)=\Delta_k(\mu)\Moore_k(\bbf)$.
\end{remark}

\begin{lemma} \label{lem:evaluation_vectors_general_k}
    Let $(\abf,\bbf)\in\Fm^t\times\Fm^\nbf$ be an evaluation pair. Let also $\lambda,\mu\in\Fm^*$, and define $(\abf',\bbf')\in\Fm^t\times\Fm^\nbf$ so that $a_i'=\lambda a_i$ for all $1\leq i \leq t$, and $b_{i,j}'=\mu b_{i,j}$ for all $1\leq i \leq t$, $1\leq j\leq n_i$. Then, $(\abf',\bbf')$ is again an evaluation pair, and $\LRS_k(\abf,\bbf)=\LRS_k(\abf',\bbf')$ for any $k\geq2$.
\end{lemma}
\begin{proof}
    By the multiplicativity of the norm and the fact that rescaling the entries of a vector does not change its rank, the pair $(\abf',\bbf')$ is still an evaluation pair. Let then $G\coloneq G_k(\abf,\bbf)$ and $G'\coloneq G_k(\abf',\bbf')$ be the standard generator matrices of $\mC\coloneq \LRS_k(\abf,\bbf)$ and $\mC'\coloneq \LRS_k(\abf',\bbf')$, respectively. By Remark~\ref{rem:diagonal_a} and Remark~\ref{rem:diagonal_mu}, for each block $1\leq i \leq t$ in $G$ and $G'$, we have
    \begin{align*}
        G_i' &= D_k(\lambda a_i) \Moore_k(\mu\bbf_i) \\
        &= D_k(\lambda) D_k(a_i)\Delta_k(\mu) \Moore_k(\bbf_i) \\
        &= \underbrace{D_k(\lambda)\Delta_k(\mu)}_{\coloneq T} \,\underbrace{D_k(a_i)\Moore_k(\bbf_i)}_{=G_i}.
        \end{align*}
        Since $T\in\GL_k(\Fm)$ and it does not depend on the considered block, we conclude that $G'=TG$, and hence $\mC=\mC'$. \qed
\end{proof}
\begin{remark} \label{rem:recovered_ev_pair}
    In the context of Problem~\ref{prob:problem_statement}, Lemma~\ref{lem:evaluation_vectors_general_k} implies that recovering an evaluation pair $(\abf,\bbf)$ with the property that $\LRS_k(\abf,\bbf)=\mC$ is not equivalent to recovering the \emph{exact} evaluation pair $(\tilde\abf,\tilde\bbf)$ that was originally used to define $\mC$. Indeed, recovering an evaluation pair $(\abf,\bbf)$ that coincides with $(\tilde\abf,\tilde\bbf)$ up to some scalar multiples $(\lambda,\mu)\in(\Fm^*)^2$ is sufficient to represent the code and to benefit from a polynomial-time decoding algorithm correcting up to half the minimum distance. As a consequence, we may assume without loss of generality that the first coordinate of $\abf$ is $1$ since we can rescale it by multiplication by $a_1^{-1}$.
\end{remark}

\subsection{Outline of the structure recoveries to follow}
In the sequel, we describe different approaches to recover an evaluation pair from the sole knowledge of a generator matrix of an LRS code. To help the reader, let us
first summarize where to find a recovery process depending on the range of parameters. 
\begin{itemize}
    \item For $2\leq k<m$, with profile $\nbf=(n,\dots,n)$ where $n\leq m$, see Section~\ref{sec:k<m};
    \item For $k=m$, with profile $\mbf=(m,\dots,m)$, except for the case $k=t=m=2$, which is of no cryptographic interest, see Section~\ref{sec:k=m};
    \item For $k=m+1$, with profile $\mbf=(m,\dots,m)$, except for the cases $t\in\{2,3\}$ and $m=2$, which remain of no cryptographic relevance, see Corollary~\ref{cor:k=m+1} in Section~\ref{sec:k>=2m};
    \item For $m+2\leq k<2m$, with profile $\mbf=(m,\dots,m)$ and
    \begin{itemize}
        \item for $t\geq3$, see Theorem~\ref{thm:full_key_recovery} in Section~\ref{sec:m<k<2m};
        \item for $t=2$, see Section~\ref{sub:t=2};
    \end{itemize}
    \item For $2m\leq k\leq tm-2$, with profile $\mbf=(m,\dots,m)$, 
    write $r=k\!\!\mod{m}$ and
    \begin{itemize}
        \item for $r\geq2$, see Theorem~\ref{thm:k>=2m} in Section~\ref{sec:k>=2m};
        \item for $r\in\{0,1\}$ and $t\geq3$, see Theorem~\ref{thm:k>=2m_r=0,1} in Section~\ref{sec:k>=2m};
    \end{itemize}
    \item The cases $k=tm-1$ and $k=tm$ are of no interest for cryptographic purposes
    and hence are not treated.
\end{itemize}

\section{Structure recovery for LRS codes with $k<n$}\label{sec:k<m}

In this section, we address Problem~\ref{prob:problem_statement} for a dimension $2\leq k< m$. In this situation, the crux of the approach consists in applying Overbeck's attack to a given block of the generator matrix.

For ease of notation, we restrict ourselves to the case in which $\mC$ has a constant profile, \textit{i.e.}, $\nbf=(n,\dots,n)\in\Z_{>0}^t$ for some $n\leq m$. Note, however, that with due care the techniques developed in this section can be adapted to the more general setting of Problem~\ref{prob:problem_statement}.

First, we show that, in the present setting, Lemma~\ref{lem:evaluation_vectors_general_k} can be strengthened to provide a complete characterization of all evaluation pairs defining the same LRS code.
\begin{lemma}\label{lem:moore_to_moore}
    Assume $k<n\leq m$. Let $\vbf,\vbf'\in\Fmn$ be two evaluation vectors and let $T\in\GL_k(\Fm)$ be such that $\Moore_k(\vbf) = T \Moore_k(\vbf')$. Then, there exists $\mu\in\Fm^*$ such that $T=\Delta_k(\mu)$. In particular, $\vbf=\mu\vbf'$.
\end{lemma}
\begin{proof}
    If $k=1$ the statement is trivial, so assume $k\geq2$.
    Let $\ebf_0, \dots, \ebf_{k-1}$ denote the canonical basis of $\Fm^k$.
    Similarly denote by $\tbf_0, \dots,\tbf_{k-1} \in \Fm^k$ the rows of $T$.
    Then for any $0\leq i \leq k-1$,
    \begin{equation}\label{eq:row_Moore}
    \ebf_i \Moore_k(\vbf) = \sigma^i(\vbf) = \tbf_i \Moore_k(\vbf').
    \end{equation} 
    For any $0 \leq i \leq k-1$ denote by $f_i \in \ore_{<k}$ the polynomial whose coefficient sequence is given by $\tbf_i$, \textit{i.e.}, $f_i(x) = \sum_{j=0} T_{i,j}x^j$ where the $T_{i,j}$ for $0 \leq i,j\leq k-1$ denote the entries of $T$.
    Thus, \eqref{eq:row_Moore} can be rephrased as
    \[
    \sigma^i(\vbf) = f_i(\sigma)(\vbf'), \quad \text{for } 0\leq i \leq k-1.
    \]
    In particular, combining the cases $i=0$ and $i=1$ yields
    \[\sigma(\vbf) = f_1(\sigma)(\vbf') = \sigma(f_0(\sigma)(\vbf')).\]
    Therefore, the polynomials $f_1(x)$ and $xf_0(x)$ are evaluate at the same $\vbf'$. Since $\deg(f_i) \leq k-1$ for all $i$, then $\deg (xf_0(x)) \leq k < n$ and, by Lemma~\ref{lem:Lagrange}, we deduce that
    $f_1(x) = x f_0(x)$. In particular $\deg(f_0) \leq k-2$. By induction, we prove that for any $1 \leq i < k$, we have $f_i = x^i f_0$ and that $\deg(f_0) = 0$. Setting $\mu \in \Fm^*$ such that $f_0 = \mu$, we deduce that $f_i = x^i \mu = \sigma^i(\mu) x^i$ for all $0 \leq i < k$. Back to the Moore matrices, this yields
    \[\sigma^i(\vbf) = \sigma^i(\mu) \sigma^i (\vbf') \quad \text{for } 0\leq i \leq k-1,\]
    hence the result. \qed
\end{proof}

\begin{proposition} \label{prop:evaluation_vectors}
    Let $2\leq k < n$ and let $(\abf,\bbf),(\abf',\bbf')\in\Fm^t\times\Fm^\nbf$ be evaluation pairs. Then, $\LRS_k(\abf,\bbf)=\LRS_k(\abf',\bbf')$ if and only if there exist $\lambda,\mu\in\Fm^*$ such that $\abf'=\lambda \abf$, and $\bbf_{i}'=\mu \bbf_{i}$ for all $1\leq i \leq t$.
\end{proposition}
\begin{proof}
    If such $\lambda,\mu\in\Fm^*$ exist, then the statement is Lemma~\ref{lem:evaluation_vectors_general_k}.
    
    Conversely, let $G\coloneq G_k(\abf,\bbf)$ and $G'\coloneq G_k(\abf',\bbf')$ respectively be standard generator matrices of $\mC\coloneq \LRS_k(\abf,\bbf)$ and $\mC'\coloneq \LRS_k(\abf',\bbf')$, and assume that $\mC=\mC'$. Then, there exists some change-of-basis matrix $T\in\GL_k(\Fm)$ such that $G'=TG$. In particular, for all $1\leq i \leq t$,
    \[D_k(a_i')\Moore_k(\bbf_i') = G_i' = TG_i = TD_k(a_i)\Moore_k(\bbf_i),\]
    that is, $\Moore_k(\bbf_i')=D_k(a_i')^{-1}TD_k(a_i)\Moore_k(\bbf_i)$. Hence, by Lemma~\ref{lem:moore_to_moore}, there exists $\mu_i\in\Fm^*$ such that $D_k(a_i')^{-1}TD_k(a_i)=\Delta_k(\mu_i)$ and $\bbf_i'=\mu_i\bbf_i$. By Remark~\ref{rem:diagonal_a} and the fact that diagonal matrices commute with each other, we get
    \[T=D_k(a_i'a_i^{-1})\Delta_k(\mu_i)\qquad\text{for every } 1\leq i \leq t.\]
    In particular, $T$ is diagonal. Since $k\geq2$, we can compare its first two diagonal entries, namely $T_{1,1}=\mu_i$ and $T_{2,2}=a_i'a_i^{-1}\sigma(\mu_i)$. Since $T$ does not depend on $i$, setting $\mu\coloneq T_{1,1}$ and $\lambda\coloneq T_{2,2}\sigma(\mu)^{-1}$ we conclude that
    \[a_i'=\lambda a_i, \quad \bbf_i'=\mu \bbf_i \quad \text{for all } 1\leq i \leq t. \tag*{\qed}\]
\end{proof}

As recalled in Section~\ref{subsec:gabidulin}, applying an Overbeck-type attack on the code defined by $G_1$ allows one to recover an evaluation vector $\bbf_1\in\Fm^n$ such that $\rowspan(G_1)=\rowspan(\Moore_k(\bbf_1))$. Since $G_1$ has full rank $k$, we can then determine the unique change-of-basis matrix $S\in\GL_k(\Fm)$ such that $SG_1=\Moore_k(\bbf_1)$, by solving a linear system.
\begin{lemma}\label{lem:recover_standard_matrix}
    Let $k<n\leq m$ and let $G=(G_1\mid \dots\mid G_t)$ be a generator matrix of an $[\nbf,k]_{q^m/q}$ LRS code. Let also $\bbf_1\in\Fm^n$ be an evaluation vector and $S\in\GL_k(\Fm)$ be such that $SG_1=\Moore_k(\bbf_1)$. Then, there exist $a_2,\dots,a_t\in\Fm^*$ and $\bbf_2,\dots,\bbf_t\in\Fm^\nbf$ such that $((1,a_2,\dots,a_t),(\bbf_1,\bbf_2,\dots,\bbf_t))$ is an evaluation pair and
    \[SG = \Big(\Moore_k(\bbf_1) \mid  D_k(a_2) \Moore_k(\bbf_2) \mid  \dots \mid  D_k(a_t) \Moore_k(\bbf_t)\Big). \]
\end{lemma}
\begin{proof}
    Since $G$ is a generator matrix of an $[\nbf,k]_{q^m/q}$ LRS code, there exist an evaluation pair $({\tilde\abf},{\tilde\bbf})\in\Fm^t\times\Fm^\nbf$ and a change-of-basis matrix $T\in\GL_k(\Fm)$ such that
    \[TG = \left(D_k({\tilde a_1})\Moore_k({\tilde\bbf_1}) \mid  D_k({\tilde a_2})
    \Moore_k({\tilde \bbf_2}) \mid  \dots \mid  D_k({\tilde a_t}) \Moore_k({\tilde
    \bbf_t})\right).\]
    In particular,
    \[\Moore_k(\bbf_1)=SG_1=ST^{-1}D_k({\tilde
    a_1})\Moore_k({\tilde\bbf_1}).\]
    Hence, by Lemma~\ref{lem:moore_to_moore}, there exists $\mu\in\Fm^*$ such that $ST^{-1}D_k({\tilde a_1})=\Delta_k(\mu)$ and $\bbf_1=\mu{\tilde\bbf_1}$. In other words,
    \[S=\Delta_k(\mu)D_k\big({\tilde a_1}^{-1}\big)T.\]
    As a consequence, by Remark~\ref{rem:diagonal_a}, for each block
    $2\leq i \leq t$ we have
    \begin{align*}
        SG_i &= \Delta_k(\mu)D_k({\tilde a_1}^{-1})D_k({\tilde a_i}) \Moore_k({\tilde\bbf_i}) \\
        &= D_k({\tilde a_1}^{-1} {\tilde a_i}) \Delta_k(\mu)\Moore_k({\tilde \bbf_i}) \\
        &= D_k({\tilde a_1}^{-1} {\tilde a_i}) \Moore_k(\mu {\tilde\bbf_i}).
    \end{align*}
    Therefore, setting $a_i\coloneq{\tilde a_1}^{-1} {\tilde a_i}$ and $\bbf_i\coloneq\mu {\tilde \bbf_i}$ for $2\leq i \leq t$, we obtain the claimed form of $SG$. In conclusion, $((1,a_2,\dots,a_t),(\bbf_1,\dots,\bbf_t))=({\tilde a_1}^{-1}{\tilde\abf},\mu{\tilde\bbf})$ is an evaluation pair by Lemma~\ref{lem:evaluation_vectors_general_k}. \qed
\end{proof}

\begin{theorem}\label{thm:key_recovery}
    Let $2\leq t < q$, $n\leq m$ and $2\leq k <n$. There exists a deterministic polynomial-time algorithm that, on input any public generator matrix $G\in\Fm^{k\times tn}$ masking an $[\nbf,k]_{q^m/q}$ LRS code $\mC$, outputs an evaluation pair $(\abf,\bbf)\in\Fm^t\times\Fm^\nbf$ such that $\LRS_k(\abf,\bbf)=\mC$.
\end{theorem}
\begin{proof}
     Write $G=(G_1\mid\dots\mid G_t)$. Since $G_1=T^{-1}D_k(\tilde a_1)\Moore_k(\tilde\bbf_1)$ for some $T\in\GL_k(\Fm)$ and some evaluation pair $(\tilde\abf,\tilde\bbf)$, the block $G_1$ generates a generalized Gabidulin code of dimension $k<n$ and has full rank $k$. Hence, an Overbeck-type attack (see Section~\ref{subsec:gabidulin}) recovers an evaluation vector $\bbf_1\in\Fmn$ such that $\rowspan(G_1)=\rowspan(\Moore_k(\bbf_1))$. Moreover, the unique matrix $S\in\GL_k(\Fm)$ such that $SG_1=\Moore_k(\bbf_1)$ is obtained by solving a linear system. By Lemma~\ref{lem:recover_standard_matrix},
     \[SG=\big(\Moore_k(\bbf_1)\mid D_k(a_2)\Moore_k(\bbf_2)\mid\dots\mid D_k(a_t)\Moore_k(\bbf_t)\big)\]
     for an evaluation pair $(\abf,\bbf)$ with $a_1=1$. For $2\leq i \leq t$, since $N_0(a_i)=1$ the first row of the $i$-th block of $SG$ is precisely $\bbf_i$. Since $\bbf_i$ is an evaluation vector, there exists an index $1\leq j\leq n$ such that $b_{i,j}\neq0$. Hence, the $(2,j)$-entry of the $i$-th block of $SG$ is $a_i\sigma(b_{i,j})$ and $a_i$ is recovered as the product of this value times $\sigma(b_{i,j})^{-1}$. This yields $(\abf,\bbf)$, and $\LRS_k(\abf,\bbf)=\rowspan(SG)=\mC$ because $S$ is invertible. Each step is a bounded number of linear-algebra operations over $\Fm$, so the procedure runs in polynomial time. A precise complexity evaluation is given in Section~\ref{subsec:complexity_k<m}.
     \qed
\end{proof}
A pseudocode description of the procedure described by Theorem~\ref{thm:key_recovery} is given in Algorithm~\ref{alg:key_recovery}.
\begin{algorithm}[ht]
\caption{Key recovery for $2\leq k< n$, $n=m$}\label{alg:key_recovery}
\begin{algorithmic}[1]
\Require $G=(G_1\mid\dots\mid G_t)\in\Fm^{k\times tn}$ generating an LRS code
\Ensure An evaluation pair $(\abf,\bbf)$ with $\LRS_k(\abf,\bbf)=\mC$
\State Find $1\leq s< m$ with $\gcd(s,m)=1$ and $\dim\Lambda_1^{(s)}(\rowspan(G_1))=k+1$
\State Compute $\Lambda_{n-k-1}^{(s)}(\rowspan(G_1))$ and recover an evaluation vector $\bbf_1$ \Comment{Overbeck}
\State Compute $T$ such that $TG_1=\Id$; $S\gets \Moore_k(\bbf_1)T$  \Comment{Gaussian elimination}
\For{$i=2,\dots,t$}
    \State $H_i\gets SG_i$
    \State $\bbf_i\gets$ first row of $H_i$; \ pick $j$ with $b_{i,j}\neq0$
    \State $a_i\gets (H_i)_{2,j}\,\sigma(b_{i,j})^{-1}$ \Comment{$(H_i)_{2,j}$: $(2,j)$-th entry of $H_i$}
\EndFor
\State \Return $((1,a_2,\dots,a_t),(\bbf_1,\dots,\bbf_t))$
\end{algorithmic}
\end{algorithm}

\subsection{Complexity}\label{subsec:complexity_k<m}
    We count operations in $\Fm$ and applications of $\sigma$. We denote by $\omega$ the complexity exponent of linear algebra.

   The Overbeck step on $\mC_1=\rowspan(G_1)$ computes the $q^s$-sums iteratively via $\Lambda_i^{(s)}(\mC_1)=\mC_1+\sigma(\Lambda_{i-1}^{(s)}(\mC_1))$. Each of the $n-k-1$ iterations applies $\sigma$ to $\mO(n^2)$ entries and performs a Gaussian elimination on a matrix with $\mO(n^2)$ rows and $n$ columns, \textit{i.e.}, $\mO(n^{\omega +1})$ operations and $\mO(n^{3})$ applications of $\sigma$ overall. Computing the dual of $\Lambda_{n-k-1}^{(s)}(\mC_1)$ and deriving $\bbf_1$ adds $\mO(n^{\omega})$ operations. Determining $S$ from $SG_1=\Moore_k(\bbf_1)$ costs $\mO(k^{\omega-1}n)$ operations, and computing $SG$ costs $\mO(k^{\omega -1}tn)$. Reading off $(\abf,\bbf)$ costs $\mO(t)$ operations and $t-1$ applications of $\sigma$. In total the procedure requires 
   \[\mO(n^{\omega+1}+n^\omega t)\ \text{operations in }\Fm \quad \text{and} \quad \mO(n^3+t) \ \text{applications of}\ \sigma.\]

\section{Structure recovery for LRS codes with $k=m$} \label{sec:k=m}
For $k \geq m$, considering a single block of the code yields the trivial code $\Fm^m$. Hence, our previous Overbeck-based approach does not extend to higher dimensional codes. In this section, we consider Problem~\ref{prob:problem_statement} in the special setting where $k=m$ and the profile of the targeted code is $\mbf\coloneq(m,\dots,m)\in\Z_{>0}^t$. We further assume that either $m\geq 3$ or $t\geq 3$. Hence, we exclude the case $k=m=t=2$, which remains of very low cryptographic interest.

In the case $k=m$, we argue directly on the systematic form of the public matrix $G$. Let $T$ be the unique matrix in $\GL_k(\Fm)$ such that
\begin{equation} \label{eq:M_i}
    TG=(\Id \mid M_2 \mid \dots \mid M_t)
\end{equation}
with $M_i\in\GL_m(\Fm)$. Let $(\tilde\abf,\tilde\bbf)$ be an (unknown) evaluation pair such that $\mC=\LRS_m(\tilde\abf,\tilde\bbf)$. By Remark~\ref{rem:recovered_ev_pair}, up to rescaling $\tilde\abf$, we may assume $\tilde a_1=1$. If $G=SG_m(\tilde\abf,\tilde\bbf)$ for some $S\in\GL_m(\Fm)$, then $T=\Moore_m(\tilde\bbf_1)^{-1}S^{-1}$. By \cite[Lemma~8]{augot2021rank}, 
\[\Moore_m(\tilde\bbf_1)^{-1}=\Moore_m({\tilde\bbf_1}^*)^{\top},\] where ${\tilde\bbf_1}^*=({\tilde b_{1,1}}^*,\dots,{\tilde b_{1,m}}^*)$ is the dual basis of $\tilde\bbf_1=(\tilde b_{1,1},\dots,\tilde b_{1,m})$ with respect to the trace map $\Tr_{\Fm/\Fq}$ (see \cite[Section~2.3]{lidl1994introduction} for more details). Hence, for
$2\leq i \leq t$, the matrices $M_i$ of \eqref{eq:M_i} satisfy
\[M_i=\Moore_m(\tilde\bbf_1^*)^{\top} D_m(\tilde a_i)\Moore_m(\tilde\bbf_i),\]
and
\begin{equation}\label{eq:Mij}
(M_i)_{j,\ell}=\sum_{h=0}^{m-1}N_h(\tilde a_i)\,\sigma^h\left(\tilde b_{1,j}^*\,\tilde b_{i,\ell}\right).
\end{equation}
\begin{lemma}\label{lem:twisted_trace}
    With the notation above, for any $2\leq i \leq t$,
    \[\tilde a_i\sigma(M_i)-M_i=\left(N_m(\tilde a_i)-1\right)(\tilde \bbf_1^*)^{\top} \, \tilde \bbf_i.\]
\end{lemma}
\begin{proof}
     Fix $j,1\leq \ell \leq m$ and set $b\coloneq\tilde b_{1,j}^*\,\tilde b_{i,\ell}$. By Remark~\ref{rem:diagonal_a} and the fact that $\tilde a_i\,\sigma\big(N_h(\tilde a_i)\big)=N_{h+1}(\tilde a_i)$ for all $h\geq0$, Equation \eqref{eq:Mij} yields
    \begin{align} \label{eq:a_sigma(M)}
        \tilde a_i\sigma\big((M_i)_{j,\ell}\big)
        &= \sum_{h=0}^{m-1} N_{h+1}(\tilde a_i)\sigma^{h+1}(b) \nonumber \\
        &= \sum_{h=1}^{m} N_h(\tilde a_i)\sigma^h(b) \nonumber \\
        &= \sum_{h=1}^{m-1} N_h(\tilde a_i)\sigma^h(b)+N_m(\tilde a_i)\,b.
    \end{align}
    Hence, subtracting ${(M_i)}_{j,\ell} = b + \sum_{h=1}^{m-1} N_h(\tilde a_i)\sigma^h(b)$ from \eqref{eq:a_sigma(M)} gives $\left(N_m(\tilde a_i)-1\right)b$, which yields the claimed identity. \qed
\end{proof}

\begin{theorem} \label{thm:k=m}
    Let either $t=2$ and $m\geq3$, or $3\leq t < q$ and $m\geq2$. There exists a deterministic polynomial-time algorithm that, on input any public generator matrix $G\in\Fm^{m\times tm}$ of an $[\mbf,m]_{q^m/q}$ LRS code $\mC$, outputs an evaluation pair $(\abf,\bbf)\in\Fm^t\times\Fm^\mbf$ such that $\LRS_m(\abf,\bbf)=\mC$.
\end{theorem}
\begin{proof}
    Let $(\tilde\abf,\tilde\bbf)$ be an unknown evaluation pair such that $\mC=\LRS_m(\tilde\abf,\tilde\bbf)$. For $2\leq i \leq t$, compute the matrix $M_i\in\GL_m(\Fm)$ as in~\eqref{eq:M_i}, and set $X_i\coloneq \sigma(M_i)M_i^{-1}$. By Lemma~\ref{lem:twisted_trace}, $\tilde a_i\sigma(M_i) = M_i+(N_m(\tilde a_i)-1)(\tilde\bbf_1^*)^{\top}\tilde\bbf_i$, hence
    \[X_i=(\tilde a_i)^{-1} \Big( \Id + \underbrace{\big(N_m(\tilde a_i)-1\big)\,(\tilde\bbf_1^*)^{\top} \, \big(\tilde\bbf_i M_i^{-1}\big)}_{\coloneq Y_i} \Big),\]
    \textit{i.e.}, $X_i$ is the sum of a scalar matrix with the rank-one term $(\tilde a_i)^{-1}Y_i$, whose column-span is $\langle(\tilde\bbf_1^*)^{\top}\rangle_{\Fm}$. Write $Y_i= (\tilde\bbf_1^*)^{\top} \, \left(\big(N_m(\tilde a_i)-1\big) \big(\tilde\bbf_i M_i^{-1}\big) \right) \coloneq \vbf_i^{\top} \ubf_i$. Then,
    \[Y_i \vbf_i^{\top} = \left( \vbf_i^{\top} \ubf_i \right) \vbf_i^{\top} = \vbf_i^{\top} \left( \ubf_i \vbf_i^{\top} \right) = \left( \ubf_i \vbf_i^{\top} \right) \vbf_i^{\top}.\]
    In particular, since $\rk(Y_i)=1$, the eigenvalues of $Y_i$ are $0$, with multiplicity $m-1$, and $\ubf_i \vbf_i^{\top}$, with multiplicity $1$. If these eigenvalues are distinct, then their corresponding eigenspaces (on the right) are $\langle \ubf_i\rangle_{\Fm}^\perp \coloneq \{\wbf\in\Fmn \st \ubf_i\wbf^{\top} = 0 \}$ and $\langle\vbf_i\rangle_{\Fm}$. Otherwise, if $\ubf_i \vbf_i^{\top}=0$, then $0$ is the unique eigenvalue with corresponding eigenspace ${\langle\ubf_i \rangle}^\top$. As a consequence, the eigenvalues of $X_i = (\tilde a_i)^{-1} \left(\Id+Y_i\right)$ are $\delta_1\coloneq (\tilde a_i)^{-1}$ and $\delta_2\coloneq(\tilde a_i)^{-1}\left( 1 + \ubf_i \vbf_i^{\top} \right)$. If $\delta_1=\delta_2$, then $\tilde a_i$ can be immediately recovered as the inverse of the unique eigenvalue of $X_i$ for all $2\leq i \leq t$. Otherwise, $\delta_1$ has multiplicity $m-1$, while $\delta_2$ has multiplicity $1$. Moreover, the eigenspace corresponding to $\delta_1$ is $\ubf_i^\perp$, while the eigenspace corresponding to $\delta_2$ is $\langle\vbf_i\rangle_{\Fm}$. If $m\geq3$, then $(\tilde a_i)^{-1}$ is the unique eigenvalue of $X_i$ of multiplicity larger than 1, so we recover the exact value $\tilde a_i$ for all $2\leq i \leq t$. Otherwise, $m=2$ and $t\geq3$, hence $X_i$ admits two distinct eigenvalues of multiplicity 1. To discriminate among them, we compute the eigenvectors of all matrices $X_2,\dots,X_t$, identify $\delta_2$ as the eigenvalue whose eigenspace is common to every block, and set $\tilde a_i\coloneq\delta_1^{-1}$.
    
    At this point, recovering some valid evaluation vectors for $\mC$ can be done as follows. For $i=2$, we can compute the matrix $a_2X_2-\Id=(\tilde\bbf_1^*)^\top \ubf_2$. Hence, picking any of its non-zero columns, we get a vector of the form $\bbf_1^*=\mu_2\tilde\bbf_1^*$ for some $\mu_2\in\Fm^*$. Let $\bbf_1$ be its dual basis, so that $\bbf_1=\mu\tilde\bbf_1$ with $\mu\coloneq(\mu_2)^{-1}$. 
    Next, take the systematic matrix $TG$ of \eqref{eq:M_i} and compute
    $\bbf_1 TG$. This vector is the unique vector of the code whose $m$ leftmost
    entries give $\bbf_1$. Hence 
    \[\bbf_1  TG = \mu \tilde{\bbf}_1 TG = \mu (~\tilde{\bbf_1} ~|~ \tilde{\bbf}_2 ~|~ \cdots ~|~ \tilde{\bbf}_t~)\]
    Hence we get $\tilde{\bbf}$ up to scalar multiplication. Since we already had $\tilde{\abf}$ up to scalar multiplication, from Lemma~\ref{lem:evaluation_vectors_general_k} we have an evaluation pair
    to describe the code.
 
    All steps in this procedure are polynomial-time. An exhaustive complexity evaluation is given in Section~\ref{subsec:complexity_k=m}. \qed
\end{proof}

A pseudocode description of the procedure described in Theorem~\ref{thm:k=m} for $t\geq3$ is given in Algorithm~\ref{alg:k=m}.
\begin{algorithm}[ht]
\caption{Key recovery for $k=m$, $t\geq3$}\label{alg:k=m}
\begin{algorithmic}[1]
    \Require Generator matrix $G\in\Fm^{m\times tm}$ of an LRS code $\mC$
    \Ensure An evaluation pair $(\abf,\bbf)$ with $\LRS_m(\abf,\bbf)=\mC$
    \State Compute $T$ such that $TG=(\Id\mid M_2\mid\dots\mid M_t)$
    \Comment{Gaussian elimination}
    \For{$2\leq i \leq t$}
        \State $X_i\gets \sigma(M_i)M_i^{-1}$
        \State Compute the eigenvalues and eigenvectors of $X_i$
    \EndFor
    \For{$2\leq i \leq t$}
        \If{$X_i$ has a unique eigenvalue}
            \State $a_i\gets$ inverse of eigenvalue of $X_i$
        \Else
            \State $a_i\gets$ inverse of the eigenvalue of $X_i$ with eigenvector $\neq\bbf_1^*$
        \EndIf
    \EndFor
    \State $\bbf_1^*\gets$ non-zero column of $a_2X_2-\Id$; $\bbf_1\gets$ dual basis of $\bbf_1^*$
    \State $\bbf \ \leftarrow \bbf_1 TG $ \Comment{We get $\bbf = (\bbf_1~|~\bbf_2~|~\cdots ~|~ \bbf_t)$}
    \State \Return $((1,a_2,\dots,a_t),(\bbf_1,\dots,\bbf_t))$
\end{algorithmic}
\end{algorithm}

\subsection{Complexity}\label{subsec:complexity_k=m}
As in Section~\ref{subsec:complexity_k<m}, we estimate the complexity in terms of operations in $\Fm$
and applications of $\sigma$.

Computing the systematic form of $G$ costs $\mO(tm^\omega)$ operations in $\Fm$, while computing the matrices $X_i=\sigma(M_i)M_i^{-1}$ for $2\leq i\leq t$ requires $\mO(tm^\omega)$ operations in $\Fm$ and $\mO(tm^2)$ applications of $\sigma$. By \cite{neiger2021deterministic}, the characteristic polynomials of $X_2,\dots, X_t$ can be computed in $\mO(t m^\omega)$ operations in $\Fm$. 

Due to the special form of these polynomials (only two eigenvalues, both in $\Fm$ and with respective multiplicities $m-1$ and $1$), the calculation of the eigenvalue is negligible. Then, the calculation of bases of eigenvectors for the $X_i$'s costs another $\mO(tm^{\omega})$ operations in $\Fm$.

Once $\bbf_1^*$ is recovered, its dual basis $\bbf_1$ and the matrices $\Moore_m(\bbf_1)M_i$ can be determined in $\mO(tm^\omega)$ operations in $\Fm$ and $\mO(tm^2)$ applications of $\sigma$. Reading off the vectors $\bbf_i$ from the first rows of these matrices costs $\mO(tm)$ operations.
    Overall, the procedure requires
\[\mO(tm^\omega)\ \text{operations in}\ \Fm \quad \text{and}\quad  \mO(tm^2)\ \text{applications of}\  \sigma.\]

\section{Structure recovery for LRS codes with $m+1\leq k < 2m$} \label{sec:m<k<2m}
In this section, we consider a special instance of Problem~\ref{prob:problem_statement} for $m+1\leq k < 2m$. Namely, we fix the profile of the targeted code to be $\mbf\coloneq(m,\dots,m)\in\Z_{>0}^t$. We also assume that $t \geq 3$, the case $t=2$ will be treated separately in Section~\ref{sub:t=2}. Note that we could have directly considered the $k>m$ case in its whole generality. However, we decided to separate this case for the sake of readability.

The strategy proposed here is to reduce to the case treated in Section~\ref{sec:k<m}. We first show that shortening $\mC$ on a full block yields an LRS code of dimension $k-m$. We then apply the procedure of Theorem~\ref{thm:key_recovery} to recover its parameters, and finally we explain how to lift the recovered parameters to a valid evaluation pair for $\mC$. \textbf{The boundary case $k=m+1$}, where shortening would not leave enough information, \textbf{is treated later in Section~\ref{sec:k>=2m}.}

\subsection{Shortening}
In this subsection, we introduce the main tool underlying the procedure developed for this range of parameters, namely, the \emph{shortening} operation. Shortening for sum-rank-metric codes was introduced in its full generality in~\cite{MartinezPenas2019Supports} and has recently been studied from a geometric perspective in~\cite{astore2026lengthmsrd}. Here, we adopt a more restrictive definition, which is sufficient for our purposes and corresponds to the special case of shortening entire blocks of the code.

\begin{definition}[Shortened code]
    Let $G=(G_1\mid\dots\mid G_t)$ be a generator matrix of $\mC\subseteq\Fm^\mbf$. For $I\subseteq\{1,\dots,t\}$, let $I^c=\{1,\dots,t\}\setminus I$, and let $G_J$ be the $k\times|J|m$ submatrix of $G$ formed by the blocks indexed by $J\subseteq\{1,\dots,t\}$. Let $K_I$ be a matrix whose rows form a basis of the left kernel of $G_I$, \textit{i.e.}, $\{\vbf\in\Fm^k\st\vbf G_I=\mathbf0\}$. We say that
    \begin{equation}\label{eq:general_shotrtened}
        \Short_I(\mC)\coloneq\rowspan(K_IG_{I^c})
    \end{equation}
    is the \emph{shortened code derived from $\mC$ with respect to the blocks indexed by $I$}.
\end{definition}
\begin{remark} \label{rem:general_shortening}
    In the setting of Problem~\ref{prob:problem_statement}, since the map $\Fm^k\rightarrow\Fmn$, $\vbf\mapsto\vbf G$ is injective, $\dim(\Short_I(\mC))=k-\rk(G_I)$. Moreover, $\Short_I(\mC)$ is the image of the subcode
    \[\mC_{(I)}\coloneq\{\cbf\in\mC\st\cbf_i=\mathbf0\text{ for all }i\in I\}\]
    under the projection $\pi_{I^c}$ onto the blocks indexed by $I^c$. In particular, $\Short_I(\mC)$ depends only on $\mC$ and $I$, but not on the choice of $G$. Moreover, the generator matrix $K_IG_{I^c}$ of $\Short_I(\mC)$ can be computed by a single Gaussian elimination and a matrix product.
\end{remark}

In this section, we will simply consider the case where $I$ is a singleton. The general case will be treated in Section~\ref{sec:k>=2m}.

We now show that $\Short_i(\mC)$ is again an LRS code, with explicit evaluation parameters.
\begin{lemma}\label{lem:evaluation_property}
    For every $\gamma\in\Fq$, the skew polynomial $x^m-\gamma$ is central in $\ore$. Moreover, for all $a,b\in\Fm$ with $N_m(a)\neq \gamma$ and $f\in\ore$,
    \[\ev_{a,b}\big((x^m-\gamma)f\big)=\big(N_m(a)-\gamma\big)\,\ev_{a,b}(f).\]
\end{lemma}
\begin{proof}
    By Remark~\ref{rem:central_xm}, for every polynomial $f=\sum_j f_jx^j$, centrality gives $(x^m-\gamma)f=\sum_j f_jx^{j+m}-\gamma\sum_j f_jx^j$. Hence, by Remark~\ref{rem:diagonal_a},
    \begin{align*}
        \ev_{a,b}\big((x^m-\gamma)f\big) &= \sum_j f_j\sigma^{j+m}(b)N_{j+m}(a) - \gamma\sum_j f_j\sigma^j(b)N_j(a) \\
        &= \big(N_m(a)-\gamma\big)\sum_j f_j\sigma^j(b)N_j(a) \\
        &= \big(N_m(a)-\gamma\big)\ev_{a,b}(f). \tag*{\qed}
    \end{align*}
\end{proof}

\begin{lemma}\label{lem:block_zero}
    Let $2\leq t < q$, $m+1\leq k < 2m$, and let $(\abf,\bbf)\in\Fm^t\times\Fm^\mbf$ be an evaluation pair. Fix $1\leq i\leq t$ and let $\cbf=(\cbf_1,\dots,\cbf_t)\in\LRS_k(\abf,\bbf)$ be a non-zero codeword such that $\cbf=\ev_{\abf,\bbf}(g)$ for $g\in\ore_{<k}$ and $\cbf_i=\mathbf 0$. Then, there exists a non-zero $f\in\ore_{<k-m}$ such that $g=(x^m-N_m(a_i))f$ and, for all $1\leq j\leq t$ with $j\neq i$,
    \[\cbf_j=\big(N_m(a_j)-N_m(a_i)\big)\big(\ev_{a_j,b_{j,1}}(f),\dots,\ev_{a_j,b_{j,m}}(f)\big).\]
\end{lemma}
\begin{proof}
    Set $\gamma_i\coloneq N_m(a_i)\in\Fq$ for the fixed index $1\leq i\leq t$. By Remark~\ref{rem:central_xm}, $x^m-\gamma_i$ belongs to the center of $\ore$. Hence, Euclidean division on the left and on the right are the same and give
    \[g = f(x^m - \gamma_i) +r = (x^m-\gamma_i)f + r,\]
    for some $f\in\ore_{<k-m}$ and $r\in\ore_{<m}$.
    In particular, $\ev_{a_i,\bbf_i}(g)=\cbf_i=0$ and Lemma~\ref{lem:evaluation_property} gives $\ev_{a_i,b_{i,\ell}}(r)=0$ for all $1\leq \ell \leq m$.
    Thus, since $\deg(r) < m$, we deduce from Lemma~\ref{lem:Lagrange} that
    \[r=0 \quad \text{and}\quad  \quad g=(x^m-\gamma_i)f.\]
    In particular, since $\cbf=\ev_{\abf,\bbf}(g)\neq0$, then $f\neq0$. Finally, by Lemma~\ref{lem:evaluation_property}, for every $1\leq j\leq t$ with $j\neq i$, $\cbf_j=(c_{j,1},\dots,c_{j,m})$ with
    \[c_{j,\ell} = \ev_{a_j,b_{j,\ell}}\big((x^m-\gamma_i)f\big) = (N_m(a_j)-\gamma_i)\ev_{a_j,b_{j,\ell}}(f), \quad 1\leq \ell\leq m. \tag*{\qed}\]
\end{proof}

\begin{proposition}\label{prop:shortened_is_lrs}
    Let $2\leq t < q$, $m+1\leq k < 2m$, and $\mC=\LRS_k(\abf,\bbf)$ for some evaluation pair $(\abf,\bbf)\in\Fm^t\times\Fm^\mbf$. Fix $1\leq i \leq t$ and set
    \begin{gather*}
        \abf'\coloneq(a_1,\dots,a_{i-1},a_{i+1},\dots,a_t) \\
        \bbf'\coloneq(\mu_1\bbf_1,\dots,\mu_{i-1}\bbf_{i-1},\mu_{i+1}\bbf_{i+1},\dots,\mu_t\bbf_t),
    \end{gather*}
    with $\mu_j\coloneq N_m(a_i)^{-1}N_m(a_j)-1$ for $1\leq j\leq t$ with $j\neq i$. Then, $(\abf',\bbf')\in\Fm^{t-1}\times\Fm^{\mbf'}$, with $\mbf'\coloneq(m,\dots,m)\in\Z_{>0}^{t-1}$, is an evaluation pair and
    \[\Short_i(\mC)=\LRS_{k-m}(\abf',\bbf').\]
\end{proposition}
\begin{proof}
    Since $\abf$ forms an evaluation pair with $\bbf$, $\mu_j\coloneq N_m(a_i)^{-1}N_m(a_j)-1\in\Fq^*$ for all $1\leq j\leq t$ with $j\neq i$. Therefore, the rank of the rescaled vectors $\bbf_j'\coloneq\mu_j\bbf_j$ is preserved, and $(\abf',\bbf')$ is still an evaluation pair.

    Let $\cbf_{(i)}\coloneq(\cbf_1,\dots,\cbf_{i-1},\cbf_{i+1},\dots,\cbf_t)\in\Short_i(\mC)$ be a non-zero codeword. By Remark~\ref{rem:general_shortening}, we have that $(\cbf_1,\dots,\cbf_{i-1},\mathbf{0},\cbf_{i+1},\dots,\cbf_t)\in\mC$. Moreover, for all $1\leq j\leq t$ with $j\neq i$, by Lemma~\ref{lem:block_zero} there exists a polynomial $f\in\ore_{<k-m}$ such that, for all $1\leq j\leq t$ with $j\neq i$,
    \[\cbf_j = \big(N_m(a_j)-N_m(a_i)\big)\ev_{a_j,\bbf_j}(f) = N_m(a_i)\,\ev_{a_j,\bbf_j'}(f) = \ev_{a_j,\bbf_j'}\left(N_m(a_i)f\right),\]
    using the $\Fq$-linearity of the map $b\mapsto\ev_{a_j,b}(f)$ and the $\Fm$-linearity of $\ev_{a_j,b}$. Therefore, $\cbf_{(i)}\in\LRS_{k-m}(\abf',\bbf')$. Hence, $\Short_i(\mC)\subseteq\LRS_{k-m}(\abf',\bbf')$. Note that the $i$-th block of $G$ corresponds to the $i$-th block of $G_k(\tilde\abf,\tilde\bbf)$, up to left multiplication by an invertible matrix. By Remark~\ref{rem:diagonal_a}, $\rk(G_i)=\rk\big(D_k(\tilde a_i)\Moore_k(\tilde\bbf_i)\big)=\rk\big(D_m(\tilde a_i)\Moore_m(\tilde\bbf_i)\big)=m$, which gives \[\dim(\Short_i(\mC))=\rk(K_i)= k-m=\dim\left(\LRS_{k-m}(\abf',\bbf')\right),\] hence we conclude that equality holds. \qed
\end{proof}

\subsection{Retrieving the evaluation parameters of the original code}\label{sub:retrieve}
We now show how to recover an evaluation pair $(\abf,\bbf)$ such that $\LRS_k(\abf,\bbf)$ coincides with $\mC=\LRS_k(\tilde\abf,\tilde\bbf)$. We first consider the generic case $t\geq3$, postponing the case of $2$-block codes to Section~\ref{sub:t=2}.

The idea is to consider two distinct shortenings of the code at a single block. In particular, we first deal with the code shortened at the first block, and then with the code shortened at the second one. For both shortened codes, we apply the procedure of Theorem~\ref{thm:key_recovery}. This is legitimate provided that the shortened code has dimension at least $2$ and length larger than its dimension, which requires $m+2 \leq k<2m$. Each run returns the parameters of all non-shortened blocks, up to a common pair of scalar multipliers and block-dependent factors. Moreover, block $3$ is left unshortened in both runs, allowing us to reconcile the two outputs and recover an evaluation pair for the original code.

\begin{theorem} \label{thm:full_key_recovery}
    Let $3\leq t < q$ and $m+2\leq k < 2m$. There exists a deterministic polynomial-time algorithm that, on input any generator matrix $G\in\Fm^{k\times tm}$ of an $[\mbf,k]_{q^m/q}$ LRS code $\mC$, outputs an evaluation pair $(\abf,\bbf)\in\Fm^t\times\Fm^\mbf$ such that $\LRS_k(\abf,\bbf)=\mC$.
\end{theorem}
\begin{proof}
    Let $(\tilde\abf,\tilde\bbf)$ be an unknown evaluation pair such that $\mC=\LRS_k(\tilde\abf,\tilde\bbf)$. For $1\leq i,j\leq t$ with $i\neq j$, set
    \[\tilde\rho_{i,j}\coloneq N_m(\tilde a_i)^{-1}N_m(\tilde a_j)-1\in\Fq^*.\]
    For $i\in\{1,2\}$, compute a generator matrix of $\Short_i(\mC)$ as in~\eqref{eq:general_shotrtened} and apply the procedure of Theorem~\ref{thm:key_recovery} to it, which is legitimate since $2\leq k-m\leq m-1$. By Proposition~\ref{prop:shortened_is_lrs}, Theorem~\ref{thm:key_recovery} and Proposition~\ref{prop:evaluation_vectors}, this returns an evaluation pair $\big(a_j^{(i)},\bbf_j^{(i)}\big)$ such that, for some $\lambda_i,\mu_i\in\Fm^*$,
    \[a_j^{(i)}=\lambda_i\tilde a_j, \quad \bbf_j^{(i)}=\mu_i\,\tilde\rho_{i,j}\,\tilde\bbf_j, \quad \text{for } 1\leq j\leq t, \ j\neq i.\]
    Since $t\geq3$, block $3$ appears in both runs, and
    \[a_3^{(1)}\big(a_3^{(2)}\big)^{-1}=\lambda_1\lambda_2^{-1}.\]
    Set $a_1\coloneq a_3^{(1)}\big(a_3^{(2)}\big)^{-1} a_1^{(2)}=\lambda_1\tilde a_1$ and $a_j\coloneq a_j^{(1)}=\lambda_1\tilde a_j$ for $2\leq j\leq t$. Then, $\abf=(a_1,\dots,a_t)=\lambda_1\tilde\abf$.

    By multiplicativity of the norm, $N_m(a_i)^{-1}N_m(a_j)=N_m(\tilde a_i)^{-1}N_m(\tilde a_j)$, hence every $\tilde\rho_{i,j}$ is computable from $\abf$. For $2\leq j\leq t$, set $\bbf_j\coloneq(\tilde\rho_{1,j})^{-1}\bbf_j^{(1)}=\mu_1\tilde\bbf_j$. Moreover, $(\tilde\rho_{2,3})^{-1}\bbf_3^{(2)}=\mu_2\tilde\bbf_3$. Since $\tilde\bbf_3$ is an evaluation vector, there exists $1\leq \ell \leq m$ such that $\tilde b_{3,\ell}\neq0$. Hence, comparing the $\ell$-th entry of $\bbf_3$ with that of $\bbf_3^{(2)}$ gives
    \[\mu\coloneq b_{3,\ell}\,\tilde\rho_{2,3}\,\big(b_{3,\ell}^{(2)}\big)^{-1}=\mu_1\mu_2^{-1}.\]
    Hence, we finally set $\bbf_1\coloneq\mu\,(\tilde\rho_{2,1})^{-1}\bbf_1^{(2)}=\mu_1\tilde\bbf_1$.

    In conclusion, by Lemma~\ref{lem:evaluation_vectors_general_k}, $(\abf,\bbf)=(\lambda_1\tilde\abf,\mu_1\tilde\bbf)$ is an evaluation pair, and $\LRS_k(\abf,\bbf)=\LRS_k(\tilde\abf,\tilde\bbf)=\mC$. All steps are polynomial-time. An exhaustive complexity evaluation will be given in the general case in Section~\ref{subsec:complexity_k>=2m}. \qed
\end{proof}

A pseudocode description of the procedure described in Theorem~\ref{thm:full_key_recovery} is given in Algorithm~\ref{alg:full_key_recovery}.
\begin{algorithm}[ht]
\caption{Key recovery for $m+2 \leq k< 2m$, $t\geq3$}\label{alg:full_key_recovery}
\begin{algorithmic}[1]
    \Require $G=(G_1\mid\dots\mid G_t)\in\Fm^{k\times tm}$ generating an LRS code $\mC$
    \Ensure An evaluation pair $(\abf,\bbf)$ with $\LRS_k(\abf,\bbf)=\mC$
    \For{$i\in\{1,2\}$}
        \State $K_i\gets$ basis of the left kernel of $G_i$ \Comment{Gaussian elimination}
        \State $(a^{(i)}_j,\bbf^{(i)}_j)_{j\neq i}\gets$ Algorithm~\ref{alg:key_recovery}$\big(K_iG_{(\hat{i})}\big)$ \Comment{$G_{(\hat{i})}\gets G$ without block $i$}
    \EndFor
    \State $\lambda\gets a^{(1)}_3/a^{(2)}_3$; \ $(a_1,\dots,a_t)\gets(\lambda a^{(2)}_1,a^{(1)}_2,\dots,a^{(1)}_t)$
    \For{$2\leq j\leq t$}
        \State $\rho_{1,j}\gets N_m(a_1)^{-1}N_m(a_j)-1$; \ $\bbf_j\gets\rho_{1,j}^{-1}\bbf^{(1)}_j$
    \EndFor
    \State Pick $\ell$ with $b_{3,\ell}\neq0$; \ $\mu\gets b_{3,\ell}\rho_{2,3}/b^{(2)}_{3,\ell}$; \ $\bbf_1\gets\mu\rho_{2,1}^{-1}\bbf^{(2)}_1$
    \State \Return $((a_1,\dots,a_t),(\bbf_1,\dots,\bbf_t))$
\end{algorithmic}
\end{algorithm}

\subsection{Case $t=2$ and $m+2\leq k < 2m$} \label{sub:t=2}
With this setting, the dual code of $\mC$ is an $[\mbf,2m-k\,]_{q^m/q}$ LRS code of dimension $2\leq 2m-k\leq m-2$. Hence, Theorem~\ref{thm:key_recovery} applies directly to $\mC^\perp$. An evaluation pair for $\mC$ is then recovered via Theorem~\ref{thm:LRS_duality} and~\cite[Section 2.5]{hormann2022distinguishing}.

Note that the only remaining case, namely $t=2$ and $k=2m-1$, corresponds to the degenerate setting in which the code $\mC$ has minimum distance $d=2m-(2m-1)+1=2$. As a consequence, its error-correcting capability is null, and we may therefore discard this case, as it is of no practical interest from a cryptographic perspective.

\section{Generalized structure recovery for LRS codes with $k\geq 2m$}\label{sec:k>=2m}

In this section, we show how to iteratively apply the procedures described in the sections above to address Problem~\ref{prob:problem_statement} for codes with $2m\leq k \leq tm$ and profile $\mbf\coloneq(m,\dots,m)\in\Z_{>0}^t$. Write $k=pm+r$ with
\[p\coloneq\left\lfloor \frac{k}{m} \right\rfloor\in\{2,\dots,t\} \quad \text{and} \quad r\coloneq k-pm\in\{0,\dots,m-1\}.\]
Following the approach of Section~\ref{sec:m<k<2m}, the main idea is to shorten several blocks at once, thereby reducing the problem to one of the configurations already treated above. We then recover the parameters of the shortened code and lift them back to the original code. However, the constructions presented in Section~\ref{sec:m<k<2m} do not directly extend to the present setting. In particular, some of the previous definitions and results, including Lemma~\ref{lem:block_zero}, need first to be generalized.

Hence, Lemma~\ref{lem:block_zero} is extended to every dimension and every set of blocks by Lemma~\ref{lem:block_zero_general} below.
\begin{lemma}\label{lem:block_zero_general}
    Assume $2\leq t < q$ and $m+1 \leq k \leq tm$. Let $(\abf,\bbf)\in\Fm^t\times\Fm^\mbf$ be an evaluation pair, $I\subseteq\{1,\dots,t\}$ with $|I|m<k$, and let $g\in\ore_{<k}$ be such that $\ev_{a_i,\bbf_i}(g)=0$ for all $i\in I$. Then, there exists a unique $f\in\ore_{<k-|I|m}$ such that
    \[g=\Big(\prod_{i\in I}\big(x^m-N_m(a_i)\big)\Big)f.\]
    Moreover, for every $j\in I^c$ and every $b\in\Fm$,
    \begin{equation} \label{eq:evaluation_prop}
        \ev_{a_j,b}(g)=\Big(\prod_{i\in I}\big(N_m(a_j)-N_m(a_i)\big)\Big)\ev_{a_j,b}(f).
    \end{equation}
\end{lemma}
\begin{proof}
    Consider the polynomial 
    \[
    h_I (x) := \prod_{i \in I} (x^m - N_m(a_i)).
    \]
    This polynomial is central in $\Fm [x; \sigma]$. Then, similarly to Lemma~\ref{lem:block_zero}, consider
    the Euclidean division of $g$ by $h_I$ giving $g = f h_I + r$ for some $r\in \Fm[x; \sigma]_{<m |I|}$.
    The vanishing constraints on $g$ together with an inductive application of Lemma~\ref{lem:Lagrange} lead to prove that $r=0$. Hence, \eqref{eq:evaluation_prop} is obtained by evaluating skew polynomials of 
    the form $g =f h_I = h_I f$ since $h_I$ is central. This yields the proof. \qed
\end{proof}

We now show that shortening on several blocks produces again
an LRS code, whose evaluation parameters depend on the product of the single-block multipliers determined in Proposition~\ref{prop:shortened_is_lrs}.
\begin{proposition}\label{prop:iterated_shortening}
    Let $2\leq t < q$, $m+1\leq k \leq tm$, and $\mC=\LRS_k(\abf,\bbf)$ for some evaluation pair $(\abf,\bbf)\in\Fm^t\times\Fm^\mbf$. Fix $I\subseteq\{1,\dots,t\}$ such that $|I|m<k$. Let $\abf'$ be the vector obtained by removing from $\abf$ the entries indexed by $I$, and set
    \[\bbf'_j\coloneq\rho_j\bbf_j, \quad \rho_j\coloneq\prod_{i\in I}\left(N_m(a_i)^{-1}N_m(a_j)-1\right)\in\Fq^*, \quad j\in I^c.\]
    Then, $(\abf',\bbf')\in\Fm^{t- |I|}\times\Fm^{\mbf'}$, with $\mbf'\coloneq(m,\dots,m)\in\Z_{>0}^{t- |I|}$, is an evaluation pair and
    \[\Short_I(\mC)=\LRS_{k-|I|m}\big(\abf',\bbf'\big).\]
    In particular, $\rk(G_I) = |I|m$.
\end{proposition}
\begin{proof}
    First of all, note that $(\abf',\bbf')$ is still an evaluation pair by Lemma~\ref{lem:evaluation_vectors_general_k}. Set \[k'\coloneq k-|I|m, \quad h_I\coloneq\prod_{i\in I}(x^m-N_m(a_i)), \quad \text{and} \quad \nu\coloneq\prod_{i\in I}N_m(a_i)\in\Fq^*,\] so that $\prod_{i\in I}(N_m(a_j)-N_m(a_i))=\nu\rho_j$ for all $j\in I^c$.

    Let $\cbf=\ev_{\abf,\bbf}(g)\in\mC_{(I)}$ for some $g\in\ore_{<k}$. By Lemma~\ref{lem:block_zero_general}, $g=h_Ig'$ for some $g'\in\ore_{<k'}$. For every $j\in I^c$, the same lemma yields
    \[\cbf_j
    =\left(\prod_{i\in I}(N_m(a_j)-N_m(a_i))\right)
    \ev_{a_j,\bbf_j}(g')\\
    =\nu\rho_j\,\ev_{a_j,\bbf_j}(g')\\
    =\ev_{a_j,\bbf_j'}(\nu g').\]
    Hence, $\pi_{I^c}(\cbf)\in\LRS_{k'}(\abf',\bbf')$.

    Conversely, let $f\in\ore_{<k'}$ and set $g\coloneq\nu^{-1}h_If\in\ore_{<k}$. Note that $\nu^{-1}\in\Fq$ is central in $\ore$. Hence, by Lemma~\ref{lem:evaluation_property}, $g$ vanishes on the blocks indexed by $I$. Moreover, by the computation above, $\ev_{\abf',\bbf'}(f)=\pi_{I^c}\left(\ev_{\abf,\bbf}(g)\right)
    \in\pi_{I^c}(\mC_{(I)})$.

    By Remark~\ref{rem:general_shortening}, $\Short_I(\mC)=\pi_{I^c}(\mC_{(I)})$. Hence, $\Short_I(\mC)=\LRS_{k'}(\abf',\bbf')$. In particular,
    \[
    \dim(\Short_I(\mC))=k-|I|m,
    \]
    and therefore, by Remark~\ref{rem:general_shortening}, $\rk(G_I)=|I|m$. \qed
\end{proof}

\begin{remark}\label{rem:uniqueness_shortened}
     Reconciling the outputs of several shortenings requires that the evaluation pair of each shortened code is determined up to a common pair $(\lambda,\mu)\in(\Fm^*)^2$, as in Lemma~\ref{lem:evaluation_vectors_general_k}. For shortened codes of dimension $2\leq k'<m$, this condition follows by Proposition~\ref{prop:evaluation_vectors}. For $k'=m$, it follows from Theorem~\ref{thm:k=m}, since the procedure described in its proof recovers $\tilde\abf$ exactly and $\tilde\bbf$ up to a scalar.
\end{remark}

\begin{definition}[$z$-covering chain] \label{def:chain}
    Let $t,z,L$ be positive integers. A sequence of sets $(I_1,\dots,I_L)$, with $I_\ell\subseteq\{1,\dots,t\}$ for $1\leq \ell\leq L$, satisfying
    \begin{gather*}
        |I_\ell|\leq z \ \text{ for } \ 1\leq \ell \leq L, \qquad \bigcap_{1\leq \ell \leq L}I_\ell=\emptyset, \\
        I_{\ell-1}\cup I_{\ell}\neq\{1,\dots,t\} \ \text{ for } \ 2\leq \ell \leq L,
    \end{gather*}
    is called a \emph{$z$-covering chain}.
\end{definition}

\begin{example} \label{ex:chain}
    In what follows, for $t$-block codes of dimension $k=pm+r<tm$ as above, we consider $p$-covering chains in which all sets have the same cardinality and $L=t-1$. More precisely, let $\bar p\coloneq p$ if $r\geq2$, and $\bar p \coloneq p-1$ otherwise. Then, $|I_j|=\bar p$ for all $1\leq j\leq L$, and consider the sets
    \[
        I_1\coloneq\{1,\dots,\bar p\}, \quad
        I_j\coloneq
        \{j,\dots,j+\bar p-1 \!\!\mod{t}\}, \quad 2\leq j\leq L.
    \]
\end{example}

\begin{lemma} \label{lem:reconciliation}
    Assume $2\leq t < q$, $2m \leq k\leq tm$ and let $\mC=\LRS_k(\tilde\abf,\tilde\bbf)\subseteq\Fm^\mbf$. Let $L$ be a positive integer and fix a $p$-covering chain $(I_1,\dots,I_L)$, where $p$ is such that $k=pm+r$ as above. Suppose further that for each $1\leq \ell \leq L$ an evaluation pair of $\Short_{I_\ell}(\mC)$ is known, of the form
    \[a_j^{(\ell)}=\lambda_\ell\,\tilde a_j, \quad \bbf_j^{(\ell)}=\mu_\ell\,\rho^{(\ell)}_j\,\tilde\bbf_j, \quad j\in I_\ell^c,\]
    with $\lambda_\ell,\mu_\ell\in\Fm^*$ and $\rho^{(\ell)}_j=\prod_{i\in I_\ell}\big(N_m(\tilde a_i)^{-1}N_m(\tilde a_j)-1\big)$. Then one recovers in polynomial time an evaluation pair $(\abf,\bbf)\in\Fm^t\times\Fm^\mbf$ with $\LRS_k(\abf,\bbf)=\mC$.
\end{lemma}
\begin{proof}
    Set $a_i \coloneq a_i^{(1)}$ for all $i\in I_1^c$. Let now $i\in I_1$, which implies $i\in\bigcup_{2\leq \ell \leq L} I_\ell^c$. Let $h:=\min\{2\leq \ell\leq L \st i\in I_\ell^c\}$. Then, $i\notin I_{h-1}^c$, that is, $i\in I_{h-1}$. Hence, $i\in I_{h-1}\cap I_h^c$. Pick $j\in I_{h-1}^c\cap I_h^c$ and set $a_i \coloneq a_{j}\big(a_{j}^{(h)}\big)^{-1}a_i^{(h)}=\lambda_1\tilde a_i$. Overall, this procedure retrieves $\abf\coloneq(a_1,\dots,a_t)=\lambda_1\tilde\abf$.

    For each $1\leq \ell \leq L$, $\rho^{(\ell)}_i$ depends only on the norms of elements in $\abf$, hence it is now computable. Set $\bbf_i \coloneq (\rho^{(1)}_i)^{-1}\bbf_i^{(1)} = \mu_1 \tilde\bbf_i$ for all $i\in I_1^c$. Let now $i\in I_1$ and $h$ be as above. Pick $j\in I_{h-1}^c\cap I_h^c$ and an index $1\leq \ell \leq m$ with $b_{j,\ell}\neq0$, and set
    \[\bbf_i \coloneq b_{j,\ell} \, \rho^{(h)}_{j} \big(b^{(h)}_{j,\ell} \rho^{(h)}_i\big)^{-1} \bbf_i^{(h)} = \mu_1 \mu_h^{-1} \big(\rho_i^{(h)}\big)^{-1} \bbf_i^{(h)} = \mu_1 \tilde\bbf_i.\] Then, we define $\bbf\coloneq(\bbf_1,\dots,\bbf_t)=\mu_1\tilde\bbf$.
    
    In conclusion, by Lemma~\ref{lem:evaluation_vectors_general_k}, $(\abf,\bbf)$ is an evaluation pair, and $\LRS_k(\abf,\bbf)=\mC$. \qed
\end{proof}

\begin{theorem}\label{thm:k>=2m}
    Let $3\leq t < q$ and $2m\leq k \leq tm-2$. Write $k=pm+r$ as above, and assume $r\geq2$. There exists a deterministic polynomial-time algorithm that, on input any generator matrix $G\in\Fm^{k\times tm}$ of an $[\mbf,k]_{q^m/q}$ LRS code $\mC$, outputs an evaluation pair $(\abf,\bbf)\in\Fm^t\times\Fm^\mbf$ such that $\LRS_k(\abf,\bbf)=\mC$.
\end{theorem}
\begin{proof}
    Assume first that $k<(t-1)m$, and write $k=pm+r$ as above. Shortening $\mC$ on any $p$ blocks produces a code of dimension $2\leq r\leq m-1$, consisting of $t-p\geq2$ unmodified blocks of $\mC$. Choose the $p$-covering chain $(I_1,\dots,I_L)$ of Example~\ref{ex:chain}. Then, for each $1\leq \ell \leq L$, compute a generator matrix of $\Short_{I_\ell}(\mC)$ as in~\eqref{eq:general_shotrtened}. By Proposition~\ref{prop:iterated_shortening}, $\Short_{I_\ell}(\mC)$ is an $r$-dimensional LRS code over $\Fm^{\mbf'}$, with $\mbf'=(m,\dots,m)\in\Z_{>0}^{t-|I_\ell|}$. Recover an evaluation pair for it by Theorem~\ref{thm:key_recovery}. By Remark~\ref{rem:uniqueness_shortened}, the recovered pair has the canonical form required by Lemma~\ref{lem:reconciliation}, and hence it returns an evaluation pair $(\abf,\bbf)\in\Fm^t\times\Fm^\mbf$ such that $\LRS_k(\abf,\bbf)=\mC$.
                
    Finally, if $(t-1)m\leq k\leq tm-2$, then $\mC^\perp$ is an $[\mbf,\,tm-k]_{q^m/q}$ LRS code of dimension $2\leq tm-k \leq m-2$. Its evaluation pair is then recovered by Theorem~\ref{thm:key_recovery}, and an evaluation pair of $\mC$ is obtained from it via Theorem~\ref{thm:LRS_duality}.

    Each step described above involves a bounded number of linear-algebra operations over $\Fm$ together with $O(t)$ base-case runs, so the procedure is polynomial. A precise complexity evaluation is given in Section~\ref{subsec:complexity_k>=2m}. \qed
\end{proof}

Before dealing with the remaining open cases not yet covered in this section, we address the final unresolved case left in Section~\ref{sec:m<k<2m}, namely, $k=m+1$.
\begin{corollary}\label{cor:k=m+1}
    Let either $t\in\{2,3\}$ and $m\geq3$, or $4\leq t < q$ and $m\geq2$. There exists a deterministic polynomial-time algorithm that, on input any generator matrix $G\in\Fm^{(m+1)\times tm}$ of an $[\mbf,m+1]_{q^m/q}$ LRS code $\mC$, outputs an evaluation pair $(\abf,\bbf)\in\Fm^t\times\Fm^\mbf$ such that $\LRS_{m+1}(\abf,\bbf)=\mC$.
\end{corollary}
\begin{proof}
    The dual code of $\mC$ has dimension $k'\coloneq(t-1)m-1$. If $t=2$ and $m\geq3$, then $k'=m-1$ and Theorem~\ref{thm:key_recovery} applies to $\mC^\perp$. Otherwise, $t-1\geq 2$ and Theorem~\ref{thm:k>=2m} applies to $\mC^\perp$. In either case, the corresponding theorem yields an evaluation pair for $\mC^\perp$, from which we recover an evaluation pair for $\mC$ via \ref{thm:LRS_duality}. Since the additional steps require only polynomial time, the overall complexity is polynomial, being dominated by the complexity of the selected key-recovery algorithm. \qed
\end{proof}

\begin{theorem}\label{thm:k>=2m_r=0,1}
    Let $3\leq t < q$ and $m\geq2$. Let also $2m\leq k < tm$. Write $k=pm+r$ as above, and assume $r\in\{0,1\}$. There exists a deterministic polynomial-time algorithm that, on input any generator matrix $G\in\Fm^{k\times tm}$ of an $[\mbf,k]_{q^m/q}$ LRS code $\mC$, outputs an evaluation pair $(\abf,\bbf)\in\Fm^t\times\Fm^\mbf$ such that $\LRS_k(\abf,\bbf)=\mC$.
\end{theorem}
\begin{proof}
    Write $k=pm+r$ as above. Shortening $\mC$ on any $p-1$ blocks produces a code of dimension $k'=m+r\in\{m,m+1\}$, containing $t-p+1\geq2$ untouched blocks of $\mC$. Choose the $(p-1)$-covering chain $(I_1,\dots,I_L)$ of Example~\ref{ex:chain}. For each $1\leq \ell \leq L$, compute a generator matrix of $\Short_{I_\ell}(\mC)$ as in~\eqref{eq:general_shotrtened}. By Proposition~\ref{prop:iterated_shortening}, $\Short_{I_\ell}(\mC)$ is a $k'$-dimensional LRS code over $\Fm^{\mbf'}$, with $\mbf'=(m,\dots,m)\in\Z_{>0}^{t-|I_\ell|}$. If $k'=m$, recover an evaluation pair for it by Theorem~\ref{thm:k=m}, otherwise if $k'=m+1$, apply Corollary~\ref{cor:k=m+1}. In both cases, the recovered pair has the canonical form required by Lemma~\ref{lem:reconciliation}, and hence it returns an evaluation pair $(\abf,\bbf)\in\Fm^t\times\Fm^\mbf$ such that $\LRS_k(\abf,\bbf)=\mC$.
        
    Each step described above involves a bounded number of linear-algebra operations over $\Fm$ together with $\mO(t)$ base-case runs, so the procedure is polynomial. A precise complexity evaluation is given in Section~\ref{subsec:complexity_k>=2m}. \qed
\end{proof}

A pseudocode description of the procedures described in Theorem~\ref{thm:k>=2m}, Corollary~\ref{cor:k=m+1}, and Theorem~\ref{thm:k>=2m_r=0,1} is given in Algorithms~\ref{alg:k>=2m} and~\ref{alg:k=m+1}.
\begin{algorithm}[ht]
\caption{Reconciliation of shortened evaluation pairs}
\label{alg:reconciliation}
\begin{algorithmic}[1]
    \Require Shortening sets $I_1,\dots,I_L$ and evaluation pairs $(\abf^{(1)},\bbf^{(1)}),\dots,(\abf^{(L)},\bbf^{(L)})$ as in Lemma~\ref{lem:reconciliation}
    \Ensure An evaluation pair $(\abf,\bbf)$ for the original code
    \State $a_i\gets a_i^{(1)}$ for all $i\in I_1^c$
    \For{$i\in I_1$}
        \State $h\gets\min\{2\leq \ell \leq L \st i\in I_\ell^c\}$
        \State Pick $j\in I_{h-1}^c\cap I_h^c$
        \State $a_i\gets a_j(a_j^{(h)})^{-1}a_i^{(h)}$
    \EndFor
    \For{$1\leq \ell \leq L$}
        \For{$j\in I_\ell^c$}
            \State $\rho_j^{(\ell)}\gets\displaystyle\prod_{i\in I_\ell}\big(N_m(a_i)^{-1}N_m(a_j)-1\big)$
        \EndFor
    \EndFor
    \State $\bbf_i\gets(\rho_i^{(1)})^{-1}\bbf_i^{(1)}$ for all $i\in I_1^c$
    \For{$i\in I_1$}
        \State $h\gets\min\{2\leq \ell \leq L \st i\in I_\ell^c\}$
        \State Pick $j\in I_{h-1}^c\cap I_h^c$ and $1\leq z \leq m$ with $b_{j,z}\neq0$
        \State $\bbf_i\gets b_{j,z}\rho_j^{(h)}(b_{j,z}^{(h)}\rho_i^{(h)})^{-1}\bbf_i^{(h)}$
    \EndFor
    \State \Return $(\abf,\bbf)$
\end{algorithmic}
\end{algorithm}

\begin{algorithm}[ht]
\caption{Key recovery for $2m\leq k< tm-1$, $t\geq3$}
\label{alg:k>=2m}
\begin{algorithmic}[1]
    \Require Generator matrix $G\in\Fm^{k\times tm}$ of an LRS code $\mC$
    \Ensure An evaluation pair $(\abf,\bbf)$ with $\LRS_k(\abf,\bbf)=\mC$
    \State $p\gets\lfloor k/m\rfloor$; \ $r\gets k-pm$
    \State \textbf{if} $r\geq2$ \textbf{then} $z\gets p$ \textbf{else} $z\gets p-1$ \textbf{end if}
    \State Choose shortening sets $I_1,\dots,I_L$ of size $z$ satisfying the hypotheses of Lemma~\ref{lem:reconciliation}
    \For{$1\leq \ell \leq L$}
        \State $I_\ell^c\gets\{1,\dots,t\}\setminus I_\ell$
        \State $K_{I_\ell}\gets$ basis of the left kernel of $G_{I_\ell}$
        \Comment{Gaussian elimination}
        \State $G^{(\ell)}\gets K_{I_\ell}G_{I_\ell^c}$
        \Comment{$G_{I_\ell^c}\gets G$ without blocks indexed by $I_\ell$}
        \If{$r=0$}
            \State $(\abf^{(\ell)},\bbf^{(\ell)})\gets$ Algorithm~\ref{alg:k=m}$(G^{(\ell)})$
        \ElsIf{$r=1$}
            \State $(\abf^{(\ell)},\bbf^{(\ell)})\gets$ Algorithm~\ref{alg:k=m+1}$(G^{(\ell)})$
        \Else
            \State $(\abf^{(\ell)},\bbf^{(\ell)})\gets$ Algorithm~\ref{alg:key_recovery}$(G^{(\ell)})$
        \EndIf
    \EndFor
    \State \Return Algorithm~\ref{alg:reconciliation}$(I_1,\dots,I_L,
    (\abf^{(1)},\bbf^{(1)}),\dots,(\abf^{(L)},\bbf^{(L)}))$
\end{algorithmic}
\end{algorithm}

\begin{algorithm}[ht]
\caption{Key recovery for $k=m+1$}
\label{alg:k=m+1}
\begin{algorithmic}[1]
    \Require Generator matrix $G\in\Fm^{(m+1)\times tm}$ of an LRS code $\mC$
    \Ensure An evaluation pair $(\abf,\bbf)$ with $\LRS_{m+1}(\abf,\bbf)=\mC$
    \State $G^\perp\gets$ basis of the right kernel of $G$
        \Comment{Gaussian elimination}
    \If{($t=2$ \myAnd $m\neq2$)}
        \State $(\abf^\perp,\bbf^\perp)\gets$ Algorithm~\ref{alg:key_recovery}$(G^\perp)$
    \ElsIf{$t\geq3$}
        \State $(\abf^\perp,\bbf^\perp)\gets$ Algorithm~\ref{alg:k>=2m}$(G^\perp)$
        \Comment{Branch $r\geq2$}
    \EndIf
    \State \Return $(\abf,\bbf)\gets$ Dual\textunderscore Ev\textunderscore Pair$(\abf^\perp,\bbf^\perp)$
    \Comment{Dual\textunderscore Ev\textunderscore Pair: \cite[Section 2.5]{hormann2022distinguishing}}
\end{algorithmic}
\end{algorithm}

\subsection{Complexity}\label{subsec:complexity_k>=2m}
    The reconciliation uses $L=\mO(t)$ shortening sets. Each set costs one Gaussian elimination $\mO(k^{\omega - 1}tm)$ to define a generator matrix of the shortened code, and one base-case recovery. This leads for $m<k<(t-1)m$ to an overall complexity of
    \[
    \mO(tm^{\omega+1}+ t^2m^\omega )\ \text{operations in}\ \Fm \quad\text{and}\quad \mO(tm^3+t^2)\ \text{applications of} \ \sigma.\] The chaining step adds $\mO(t^2)$ norm evaluations and scalar operations which are negligible compared to the other calculations. The case of $k\geq (t-1)m$ requires the calculation of the dual, which is $\mO({(mt)}^{\omega})$ operations in $\Fm$ and an application of Theorem~\ref{thm:LRS_duality} which is $\mO(tm^2)$ operations in $\Fm$ and $\mO(tm^2)$ applications of $\sigma$ and finally an application of Theorem~\ref{thm:key_recovery} yields
    an additional $\mO(m^{\omega+1}+ m^{\omega}t)$ operations in $\Fm$ and $\mO(m^3+t)$ applications of $\sigma$.
    Overall, for $k \geq (t-1)m$, the procedure costs
        \[\mO(m^{\omega+1} + (tm)^{\omega})\ \text{operations in}\ \Fm
        \quad \text{and}\quad \mO(m^3+tm^2)\ \text{applications of}\ \sigma.\]

\subsection{Cases $k=tm-1$ and $k=tm$.} \label{subsec:upper_cases}
Similarly to what we observed in Section~\ref{sec:m<k<2m} for $t=2$, these cases correspond to degenerate settings in which the code $\mC$ has minimum distance $d=tm-k+1\leq2$. As a consequence, the error-correcting capability, given by $\lfloor (d-1)/2 \rfloor$, is zero in both cases. Hence, such codes are of limited interest from a cryptographic perspective and are therefore discarded. In particular, these scenarios include the cases $t=m=2$ and $k=2m-1=m+1$ or $k=2m$, which are left out by Corollary~\ref{cor:k=m+1} and Theorem~\ref{thm:k>=2m_r=0,1}, respectively.

\section{An alternative approach for recovery and possible generalization to arbitrary profiles}

\subsection{Using the Overbeck-like reconstruction from H\"ormann, Bartz and Horlemann}\label{subsec:using_Trautmann}
The previous sections described processes that permit to recover simultaneously
$\abf$ and $\bbf$. However, it was proved in \cite[Theorem~6]{hormann2022distinguishing}
that, given the knowledge of $\abf$ one can deduce that of $\bbf$ using
an Overbeck-like approach. Their recovery process is claimed to have a complexity $\mO(N^5)$ where $N$ denotes the code length.

Therefore, it should be emphasized that our shortening approach could have been used only to recover the $a_i$'s and then run the technique of \cite[Theorem~6]{hormann2022distinguishing} only at the cost of a slightly higher complexity.

\subsection{About general profiles}
In most sections above, our work restricted to the case of codes with profile $\mbf=(m,\dots, m)$. However, the general setting presented in Problem~\ref{prob:problem_statement} can probably be treated analogously. Let us sketch the general ideas.
To limit the amount of tedious notation, we limit to the case of a constant profile $\nbf=(n,\dots, n)$ with $n < m$. 

For this more general profile, the shortening operation presented in Section~\ref{sec:m<k<2m} still provides an LRS code with mostly the same $\abf$, while $\bbf$ gets modified as explained the following statement. To ease the discourse, we first discuss the shortening operation for a single block.
\begin{proposition}\label{prop:short_general profile}
Let $\mC = \LRS_k(\abf, \bbf)$ and $k \geq n$. Let $1\leq i \leq t$, then
\[\Short_i(\mC) = \LRS_{k-n}(\abf', \bbf'),\]
where $(\abf', \bbf')$ is an evaluation pair and $\abf'=(a_1, \dots, a_{i-1},a_{i+1}, \dots, a_t)$.
\end{proposition}
\begin{proof}
By Lemma~\ref{lem:Lagrange}, there exists $\varphi_i \in \Fm [x; \sigma]_{\leq n}$ such that $\ev_{a_, \bbf_i}(\varphi_i) = 0$.
For $f\in \Fm[x; \sigma]_{<k}$ we have a right Euclidean division $f = g\varphi_i + r$
with $\deg r < n$. If $\ev_{a_i, \bbf_i}(f) = 0$, then we deduce that $\ev_{a_i, \bbf_i}(r) = 0$. Then the degree bound $\deg r < n$ together with Lemma~\ref{lem:Lagrange} entail that $r = 0$.

Consequently, any $f\in \Fm[x; \sigma]_{<k}$ satisfying $\ev_{a_i, \bbf_i}(f) = 0$,
\emph{i.e.}, any $f$ giving rise to a codeword of the shortened code at block $i$, is of the form $f = g \varphi_i$ for some $g \in \Fm [x; \sigma]_{<k-n}$.
Next for any $j \neq i$,
\begin{align*}
    \ev_{a_j, \bbf_j}(f) = \ev_{a_j, \bbf_j}(g \varphi_i) &= (g(a_j \sigma)\circ\varphi_i(a_j\sigma))(\bbf_j)\\
    & = (g(a\sigma))(\varphi_i(a \sigma(\bbf_j)))\\
    &= \ev_{a_j, \ev_{a_j, \bbf_j}(\varphi_i)}(g). \tag*{\qed}
\end{align*}
\end{proof}

The conclusion of the previous statement is that if $n<m$ the shortening of 
a block may affect the shape of the $\bbf_j'$ of the non shortened blocks, while 
\textbf{the $a_j$'s of the non-shortened blocks remain unchanged}. By induction, 
when shortening more than one block, the non-shortened $a_j$'s still remain unchanged.

This leads to a recovery process that we sketch below. Let $\mC = \LRS_{k}(\abf, \bbf')$. Since the case $k< n$ was already treated in Section~\ref{sec:k<m}, we assume $k \geq n$.
\begin{enumerate}
    \item If $k > tn/2$, then switch to the dual code.
    \item From Lemma~\ref{lem:evaluation_vectors_general_k}, one can assume $a_1 = 1$. By Proposition~\ref{prop:short_general profile}, shortening gives rise to another LRS code with the same $a_i$'s. Moreover, we may shorten enough blocks at once, without touching the first block, to obtain a code of dimension $< n$. Then, from Lemma~\ref{lem:evaluation_vectors_general_k}
    the obtained code has a unique $\abf$ such that $a_1 = 1$. Hence, we uniquely determine $a_i$ for the other non shortened blocks. Note that since we assume $k < nt/2$, there remains at least $2$ non-shortened blocks. Therefore, iterating this process permits to show that, fixing $a_1 = 1$, the other $a_i$'s are uniquely determined and can be computed by first shortening, and then applying the techniques of Sections~\ref{sec:k<m} or~\ref{sec:k=m}.
    \item Once, $\abf$ is computed, run the Overbeck-like attack of \cite[Section~5.2]{hormann2022distinguishing} to recover $\bbf$.
\end{enumerate}

\section*{Conclusion}
In this paper, we gave a polynomial-time full structure recovery strategy for LRS codes with profile $(m, \dots, m)$, and sketched a recovery process for arbitrary constant profiles. It is, however, likely that the technique developed in this work extend to arbitrary profiles, at the cost of more technical attack descriptions. In summary, our results show that LRS codes cannot be used as they are for McEliece-like encryption.

\section*{Declaration of Generative AI and AI-assisted Technologies in the Writing Process}
The authors used AI-assisted tools exclusively for language editing and to improve the clarity and readability of the manuscript. All mathematical results, proofs, and scientific conclusions are the authors' own. The authors have reviewed the entire manuscript and take full responsibility for its content.

\bibliographystyle{splncs04}
\bibliography{biblio.bib}

\end{document}